\documentclass[11.5pt,letterpaper]{article}

\usepackage{mathpazo}
\usepackage{verbatim}
\usepackage{float}
\usepackage{array}
\usepackage{booktabs}
\usepackage{amsmath}
\usepackage{mathtools}
\usepackage{amssymb}
\usepackage{graphicx}
\usepackage{setspace}
\usepackage{natbib,epstopdf}
\usepackage{algorithm}
\usepackage{algorithmic}
\usepackage{xcolor, colortbl}
\usepackage{pdflscape}
\usepackage{hyperref}
\usepackage{multirow}
\usepackage{lscape}
\usepackage{xcolor, soul}
\usepackage{appendix}
\usepackage{longtable}
\usepackage{multirow}
\usepackage{rotating}
\usepackage{bm}
\sethlcolor{yellow}
\usepackage{subfig}
\usepackage{enumitem}
\usepackage{placeins}
\usepackage{longtable}
\usepackage{threeparttable}

\usepackage[utf8]{inputenc} 
\usepackage[T1]{fontenc}    
\usepackage{xspace}

\usepackage{pifont} 

\definecolor{lightgray}{gray}{0.9}
\definecolor{myr}{rgb}{0.6,0,0}
\definecolor{myb}{rgb}{0,0,0.6}
\definecolor{myg}{rgb}{0,0.4,0}
\hypersetup{colorlinks,
    citecolor=myb,
    filecolor=myb,
    linkcolor=myr,
     urlcolor=myg}

\emergencystretch=\maxdimen
\bibpunct{(}{)}{;}{a}{,}{,}

\definecolor{revblue}{rgb}{0,0,0.85}

\newcommand{\method}{{\scshape{MacroCast}}\xspace}

\newcommand{\tsfm}{{\scshape{TSFM}}\xspace}

\newcommand{\chronos}{{Chronos}\xspace}

\makeatletter
\let\orig@includegraphics\includegraphics
\renewcommand{\includegraphics}[2][]{%
  \IfFileExists{#2}%
    {\orig@includegraphics[#1]{#2}}%
    {\fbox{\parbox[c][2.5cm][c]{0.9\linewidth}{\centering\ttfamily\small
       [missing figure]\\\detokenize{#2}}}}%
}
\makeatother

\makeatletter
\let\orig@input\input
\renewcommand{\input}[1]{%
  \IfFileExists{#1}{\orig@input{#1}}{%
   \IfFileExists{#1.tex}{\orig@input{#1.tex}}{%
     \par\noindent\fbox{\ttfamily\small [missing table: \detokenize{#1}]}\par}}%
}
\makeatother

\begin{document}
\title{\method: A Vintage-Consistent Time Series Foundation Model for Real-Time Macroeconomic Forecasting\thanks{The views and opinions expressed here are those of the authors and do not necessarily reflect the views or positions of any entities they are affiliated with.}}

\author{\Large Andrea Carriero\thanks{
School of Economics and Finance, Queen Mary University of London. Email:
\href{mailto:a.carriero@qmul.ac.uk}{a.carriero@qmul.ac.uk}} , 
\ Davide Pettenuzzo\thanks{
School of Business and Economics, Brandeis University. Email:
\href{mailto:dpettenu@brandeis.edu}{dpettenu@brandeis.edu}} , 
\ and Shubhranshu Shekhar\thanks{
School of Business and Economics, Brandeis University. Email:
\href{mailto:sshekhar@brandeis.edu}{sshekhar@brandeis.edu}} }
\date{\today}
\maketitle
\begin{abstract}

\noindent We introduce \method, a lightweight Time Series Foundation Model (TSFM) for real-time macroeconomic forecasting. Existing TSFMs suffer from data leakage in two forms: temporal contamination, as the model may have seen the realized values of the series it forecasts, and revision bias, as training on fully revised data diverges from the preliminary, vintage-specific releases available to real-time forecasters. \method is, to our knowledge, the first TSFM that rules out both forms of leakage entirely: at no stage of training is the model exposed to information that would not have been available to a forecaster in real time. We train \method first on purely synthetic time series in approximately one GPU-day and then fine-tune it on synthetic time series drawn from Bayesian VARs, dynamic factor models, and ARIMA specifications estimated on vintage-specific ALFRED data. Because pretraining uses only simulated data and fine-tuning uses only real-time vintages, no observed future or revised value ever enters the model; each fine-tuning run takes nine minutes. Evaluated on the FRED-MD database in a genuine real-time out-of-sample exercise, \method improves on the AR(1) benchmark for roughly 80\% of series--horizon pairs, matches or surpasses Chronos-2 --- the strongest currently available TSFM --- and outperforms the Bayesian VAR and dynamic factor model benchmarks, all in a data-leakage-free manner.

\bigskip


\bigskip


\end{abstract}

\thispagestyle{empty}

\doublespacing

\section{Introduction}
\label{sec:intro}

The past few years have witnessed a surge of interest in Time Series Foundation Models (TSFMs), i.e. large-scale pretrained models designed to forecast across diverse domains without the need for task-specific training. This surge has been inspired by the enormous success of foundation models in natural language processing and computer vision, which has invigorated researchers at both academic and private institutions to try to adapt the same paradigm to time series data. Prominent examples include TimeGPT \citep{garza2023timegpt1}, Lag-Llama \citep{rasul2023lag}, Chronos \citep{ansari2024chronos} and its more recent successor Chronos-2~\citep{ansari2025chronos}, MOMENT \citep{goswami2024moment}, TimesFM \citep{das2024decoder}, Moirai \citep{woo2024unified} and its follow-up Moirai-2.0 \citep{liu2025moirai}, Tiny Time Mixers \citep{ekambaram2024ttms}, TEMPO \citep{cao2024tempopromptbasedgenerativepretrained}, and Time-LLM \citep{jin2023time}, among many others. Comprehensive surveys of this rapidly evolving landscape are provided by \cite{Liang_etal_2024} and \cite{wen2022transformers}. The appeal of these models is clear: by pretraining on vast and heterogeneous collections of time series, TSFMs hold the promise of a practical off-the-shelf tool for forecasters and practitioners alike across many domains.

Despite this promise, adopting TSFMs for real-time forecasting poses unique challenges. First, these models are expensive to train: they require massive datasets, substantial computational resources, and long training runs, placing them out of reach for most academic and policy-oriented forecasters.\footnote{For example, TimesFM \citep{das2024decoder} has 200 million parameters and was pretrained on approximately 100 billion observations; Chronos \citep{ansari2024chronos} scales to 710 million parameters trained on over 84,000 time series; Moirai-MoE \citep{liu2024moiraimoe} reaches 311 million parameters.} Second, existing TSFMs are susceptible to data leakage, which in the forecasting context takes two distinct forms. The first is temporal contamination: a model trained on data available up to some recent cutoff date may have seen the actual historical realizations of the very series it is later asked to forecast, invalidating any pseudo out-of-sample exercise \citep{Meyer:etal:2025:leakage, Carriero:Pettenuzzo:Shekhar:2025:LLM}. This form of leakage is a general concern that can affect any TSFM regardless of application domain. The second form is more subtle and inherently domain-specific and can be easily illustrated using the example of macroeconomic data. These data are subject to substantial and systematic revision after their initial release: for instance, a first estimate of GDP growth is published within weeks of the reference quarter and then revised repeatedly over the following quarters as more complete source data become available, with mean absolute revisions exceeding one percentage point at an annualized rate \citep{Croushore:2011}. The consequences of these revisions for empirical inference have been extensively documented in the macroeconomic forecasting literature \citep{Croushore:Stark:2001, Orphanides:2001}. \citet{Ghysels:etal:WP:2014}, for instance, find that the use of real-time rather than fully revised data substantially reduces the predictive power of macroeconomic variables for future bond returns, as well as the implied countercyclicality of term premiums. A model trained on fully revised historical values has therefore implicitly learned from information --- the final, refined readings of GDP growth, employment, or inflation --- that would not have been available to a real-time forecaster at the time of the forecast. 

The interaction between expensive training cost and data leakage creates a genuine tension. One natural response to the leakage problem is to train TSFMs "in real time", that is, using exclusively information available at the time the forecast is made. However, training or retraining existing TSFMs in real time is prohibitively expensive given the computational demands of these large models. 

This paper proposes a lightweight approach designed to resolve both types of leakage at once. We introduce \textbf{\method}, a new TSFM that builds on the TempoPFN architecture \citep{tempopfn2026} but is designed from the ground up for real-time macroeconomic forecasting. Our model is comparatively much smaller, with only one million parameters --- one to two orders of magnitude fewer than the foundation models we benchmark against (Moirai2-Small, with roughly 10 million parameters, and Chronos-2, with 120 million) and up to three orders of magnitude fewer than the largest TSFMs in the literature (TimesFM at 200 million, Moirai-MoE at 311 million, and Chronos at 710 million) --- making it extremely fast to train and fine-tune. 

Training proceeds in two phases. In the first, general phase, we pretrain \method exclusively on purely synthetic time series, so that no observed data enters the pretraining stage and temporal contamination is ruled out by construction; this step takes approximately one day on a single GPU. In the second, real-time phase, we fine-tune the model using actual macroeconomic data, but importantly we do so in a vintage-consistent manner, that is, we use vintage-specific data from the St.~Louis Fed's ALFRED repository, i.e. the actual data releases that would have been available to a forecaster at each point in time. 

This vintage-consistent fine-tuning adapts \method to the features of macroeconomic series while fully respecting the time consistency and revision history of the data, so that the fine-tuning stage, like the synthetic pretraining stage, never exposes the model to information that was unavailable to a forecaster in real time, making the model ready for proper real-time use. Crucially, each fine-tuning run takes only nine minutes, making it entirely feasible to retrain the model with each new data vintage. 

Furthermore, our fine-tuning is tailor-made for macroeconomic forecasting in that it is grounded in state-of-the-art macroeconometric domain knowledge. As is customary in foundation models, at the fine-tuning stage we expand the available data through synthetic data augmentation \citep{iglesias2023data, darlow2023tsmix}, mirroring the approach adopted by several prominent TSFMs \citep{ansari2024chronos}; unlike them, however, our synthetic data augmentation is built using a suite of well-established econometric models, each chosen to reproduce a specific, well-documented feature of macroeconomic data: Bayesian VARs \citep{Litterman1986, Banburaetal2010} for cross-variable lead--lag dynamics, dynamic factor models \citep{Stock:Watson:JASA:2002} for factor co-movement, as well as univariate ARIMA specifications.

Note that while fine-tuning on domain-specific real-time data is in principle possible also in traditional transformer-based TSFMs, it is practically very hard to implement for large models: fine-tuning a TSFM with hundreds of millions of parameters requires substantial computational resources, and doing so at the frequency required for real-time macroeconomic monitoring would be infeasible for most institutions. Moreover, contributions such as \citet{Carriero:Pettenuzzo:Shekhar:2025:LLM} and \citet{Guerron:Kumar:Liang:2026:TSFM} found that the benefits from fine-tuning large transformer models with macroeconomic data are limited.\footnote{\citet{Guerron:Kumar:Liang:2026:TSFM} find that fine-tuning Chronos-2 --- a 120-million-parameter model --- on a short macroeconomic sample actually \emph{underperforms} the zero-shot baseline across most variable--horizon combinations, a result they attribute to overfitting to the estimation-sample dynamics at the cost of the broader inductive biases acquired during pretraining on a large and diverse corpus. A similar finding is documented in \citet{Carriero:Pettenuzzo:Shekhar:2025:LLM}, where fine-tuning large \tsfm{}s directly on macroeconomic series delivers little improvement over zero-shot application. } 

In a genuine real-time out-of-sample exercise on the FRED-MD database, \method{} improves on the AR(1) benchmark for roughly 80\% of series--horizon pairs (86\% at the one-month horizon), matches or surpasses Chronos-2 --- the strongest currently available TSFM, which it beats in the majority of head-to-head comparisons --- and outperforms the Bayesian VAR and dynamic factor model benchmarks, all while remaining free of both forms of leakage.

\method{} contributes to a rapidly growing literature that adapts \tsfm{}s to macroeconomic forecasting. Because economic and financial series account for only a small fraction of the training data in general-purpose \tsfm{}s \citep{Koyuncu:etal:2026:BISTRO}, their transferability to macroeconomic settings is not guaranteed. \citet{Carriero:Pettenuzzo:Shekhar:2025:LLM} provide a systematic comparison of several \tsfm{}s against traditional macro forecasting methods on the FRED-MD database, finding that a hybrid approach combining econometric structure with \tsfm{} adaptability delivers consistent improvements over either class alone --- a finding that directly motivates the vintage-consistent fine-tuning design of \method. \citet{FL2024} show that conditional LLM-based inflation forecasts achieve lower mean-squared errors than the Survey of Professional Forecasters at most horizons. The Bank for International Settlements recently introduced BISTRO \citep{Koyuncu:etal:2026:BISTRO}, a transformer fine-tuned on the BIS macroeconomic repository, which correctly anticipates the persistence of the 2021 inflation surge. Most recently, \citet{Guerron:Kumar:Liang:2026:TSFM} benchmark Chronos-2 against BVAR across small, medium, and large variable sets from FRED-MD, documenting that Chronos-2 outperforms BVAR on average with its advantage most pronounced during crisis periods when structural breaks render linear models unreliable; they further extend structural identification and scenario analysis to the foundation-model setting. Unlike \method, however, none of these approaches simultaneously eliminates temporal contamination and revision bias: they either apply large pretrained models zero-shot, fine-tune on observed (fully revised) data, or both.

The remainder of the paper is organized as follows. Section~\ref{sec:model} describes the \method model, covering the architecture, the synthetic pretraining procedure, and the vintage-consistent fine-tuning strategy. Section~\ref{sec:empirical} presents the empirical evaluation, including the data, benchmark competitors, and forecast results. Section~\ref{sec:conclusion} concludes.

\section{\method}
\label{sec:model}

This section introduces \method. While most existing \tsfm{}s are built on the self-attention mechanism of \citet{vaswani2017attention}, we depart from this paradigm and instead build \method around a linear Recurrent Neural Network (RNN) architecture. Training proceeds in two steps, pretraining and fine-tuning. In the first, the model is pretrained on approximately ten million synthetic time series drawn from a diverse set of data-generating processes, producing a compact one-million-parameter TSFM with no exposure to any observed macroeconomic data. In the second, this pretrained model is fine-tuned on synthetic datasets calibrated to real-time FRED-MD vintages, adapting it to the specific features of macroeconomic data.  We detail the model architecture and estimation of the model in Section~\ref{sec:architecture} and Section~\ref{sec:estimation_obj}, the synthetic pretraining procedure in Section~\ref{sec:pretraining}, and the vintage-consistent fine-tuning strategy in Section~\ref{sec:finetuning}.

\subsection{Model architecture}
\label{sec:architecture}
Most existing \tsfm{}s are built on the self-attention mechanism of \citet{vaswani2017attention}, which allows each observation in a time series to be directly informed by every other observation in that series, at any lag distance. This flexibility allows the model to learn arbitrarily complex cross-lag dependence patterns, but comes at a steep computational price: the mechanism has $\mathcal{O}(T^2)$ complexity in both computation and memory ($T$ being the length of the input sequence), so cost grows quadratically with series length. Models such as \chronos \citep{ansari2024chronos}, TimesFM \citep{das2024decoder}, and Moirai \citep{woo2024unified} manage this by grouping consecutive observations into non-overlapping patches of size $P$, reducing the effective sequence length to $T/P$ and applying attention at the patch level. Patching is an efficient compromise, but may still introduce unintended effects. By construction, patching discards all variation occurring within a patch, including potentially informative within-patch dynamics and cross-series co-movements at frequencies finer than the patch size. For macroeconomic forecasting, where the precise timing and sequence of month-to-month movements (a turning point in output growth or a sudden acceleration in inflation) often carries meaningful predictive content, this loss of high-frequency detail can be consequential. Moreover, the cost of re-estimating a large transformer-based model remains substantial, and updating such a model at every forecast vintage is computationally prohibitive for most institutions.

We build \method on the TempoPFN architecture \citep{tempopfn2026}, which replaces self-attention with a linear Recurrent Neural Network (RNN). RNNs have a well-established track record as forecasting tools, and unlike traditional time series methods such as ARMA and VAR models which parametrize temporal dependence through a fixed number of lags, RNNs summarize the relevant history in a latent \emph{state} vector that is updated recursively as each new observation arrives, similar to how the state vector in a linear state-space model operates \citep{Hewamalage:etal:2021:RNN}. This feature lets an RNN condition on information that extends beyond any pre-specified lag order. The empirical potential of these tools was confirmed when an RNN-based method won the M4 competition \citep{Smyl:2020:M4}, a large-scale evaluation spanning 100,000 series across multiple domains and frequencies \citep{Makridakis:etal:2020:M4}. Furthermore, \citet{Hewamalage:etal:2021:RNN} show that, when properly configured, RNNs can be very competitive against ARIMA and ETS benchmarks also in multivariate settings, as they can pool information across related series.

\subsubsection{Recurrent neural network}
An RNN can be understood as a state-space model in which the latent state evolves according to transition equations that need not be linear or Gaussian, as \citet{Rangapuram:etal:2018:DSSM} formalize. We build \method by relying on a linear RNN, a computationally efficient variant in which the hidden-state recurrence is restricted to be linear,  while nonlinear projections at the input and output stages preserve the model's ability to capture complex dynamics. We make this connection precise below.
\begin{align}
	\bm{y}_t &= g(\bm{H}_t,\, \bm{x}_t), \label{eq:linear_rnn}\\
	\bm{H}_t &= \bm{A}(\bm{x}_t)\,\bm{H}_{t-1} + \bm{B}(\bm{x}_t)\,\bm{x}_t,\label{eq:linear_rnn2}
\end{align}
where $\bm{x}_t$ is the vector of observed inputs at time $t$, $\bm{H}_t$ is the latent state vector summarizing the history up to time $t$, $\bm{A}(\bm{x}_t)$ is the input-dependent state-transition matrix, $\bm{B}(\bm{x}_t)$ maps inputs into the state space, and $g(\cdot)$ is the observation equation mapping the current state and input to the model output $\bm{y}_t$. Two features distinguish this system from a textbook linear state-space model. First, the equations carry no explicit error term: the recurrence is deterministic given the inputs, and stochasticity enters only through the conditional quantiles produced by $g(\cdot)$, as described in Section~\ref{sec:estimation_obj}. Second, the observed inputs $\bm{x}_t$ enter the state transition itself through the input-dependent matrices $\bm{A}(\bm{x}_t)$ and $\bm{B}(\bm{x}_t)$; the recurrence is therefore linear in the state $\bm{H}_{t-1}$ but nonlinear in the inputs $\bm{x}_t$.  The full sequence of states can be computed in parallel rather than period by period \citep{grazzi2025unlocking}, reducing computational cost from $\mathcal{O}(T^2)$ to $\mathcal{O}(T)$ and allowing linear RNNs to be trained at the same speed as transformers, without sacrificing the state-space structure that makes the model interpretable.

The latent state $\bm{H}_t$ plays the role of the model's memory: it is a fixed-dimensional matrix that compresses all information in the observed history up to time $t$ into a compact representation from which forecasts are produced via the observation equation $g(\cdot)$. The practical implication of this fixed-dimensional memory is that the model cannot retrieve individual past observations directly when forming a forecast; instead, all historical information must pass through $\bm{H}_t$. In the context of macroeconomic data, however, this constraint does not turn out to be binding in practice, as the typical context in our evaluation spans up to ten years of monthly observations, well within the range where a well-designed linear RNN retains the information needed for accurate multi-step forecasts.\footnote{What the TSFM literature calls the \emph{context window}---the span of past observations the model conditions on when forming a forecast---is the direct analogue of the estimation, or in-sample, window in econometrics; we use the term context window throughout to stay close to the TSFM terminology.} The efficiency gain, on the other hand, is decisive. With linear recurrence the fine-tuning step that adapts \method to a new vintage completes in nine minutes, making vintage-consistent rolling estimation feasible over four decades of history without specialized hardware. 

Returning to the recurrence in equations~\eqref{eq:linear_rnn}-\eqref{eq:linear_rnn2}, the specific recurrence in \method is the GatedDeltaProduct introduced by \cite{siems2026deltaproduct}. A single application of the state update, advancing the state from $\bm{H}_{t-1}$ to $\bm{H}_t$, constitutes the recurrence, and a model ``layer'' is one such recurrent block together with the additional operations described below. \method stacks three such layers. The key design choice is that the state-transition matrix $\bm{A}(\bm{x}_t)$ is full, rather than restricted to a fixed diagonal structure as in simpler recurrent models. This richer parameterization gives the model greater capacity to selectively incorporate new observations into $\bm{H}_t$ while preserving relevant features of the past. \cite{grazzi2025unlocking} show that this property matters empirically for tasks where accurately tracking earlier observations is important for prediction. For macroeconomic series, where most of the predictive signal sits in recent lags, we expect this added capacity to matter less through the tracking of distant observations per se than through its ability to represent nonlinear and recurring patterns. Each model layer augments this recurrence with a short moving-average filter over adjacent observations (capturing immediate short-run dynamics), a normalization step for training stability, and a nonlinear transformation that adds flexibility beyond what the recurrence alone provides.

One subtlety of applying a recurrent model to forecasting is that observations are processed in chronological order: the model reads the data one period at a time, updating $\bm{H}_t$ as it progresses through the observed history. By the time it reaches the last observed period $T$, it holds a single state vector $\bm{H}_T$ summarizing everything it has seen. While $\bm{H}_T$ already compresses the full observed history (an advantage over fixed-lag methods) all forecast horizons would nonetheless condition on the \emph{same} state. \method addresses this through \emph{state weaving} \citep{tempopfn2026}: the model alternates between processing observed history and forecast-horizon stages, using its own predicted outputs as inputs during forecast stages in place of unobserved future realizations. In econometric terms this is closely related to iterating a dynamic model forward: multi-step forecasts are produced recursively by feeding the model's own one-step predictions back in as inputs, exactly as one iterates a VAR or a state-space model beyond the one-step-ahead horizon. The hidden state at the end of each stage becomes the starting point for the next, so each successive stage begins from an increasingly refined summary of the observed history. By the final stage, the forecasts effectively condition on a more refined historical signal, without requiring additional parameters or a second pass through the data.

\subsection{Model Inputs and Training Objective}
\label{sec:estimation_obj}

Before entering the model, each input series is standardized. This step helps with numerical stability, since after standardization the model operates in a comparable numerical range regardless of the level or scale of the underlying series. In addition, missing observations are replaced by a dedicated, learnable missing-value embedding, allowing the model to accommodate unbalanced panels without a separate imputation step.

Forecasts take the form of conditional quantiles. For each series $i = 1, \ldots, N$ and horizon $h = 1, \ldots, H$, the model produces nine quantile estimates $\hat{y}_{i,T+h}^{(\tau)}$ for levels $\mathcal{Q} = \{0.1, 0.2, \ldots, 0.9\}$ by applying a series-specific linear transformation to the hidden state $\bm{H}_{T+h}$ from the final processing stage. The model is estimated by minimizing the pinball loss aggregated across all series, quantile levels, and forecast horizons,
\begin{align}
	\mathcal{L} = \sum_{i=1}^{N}\sum_{\tau \in \mathcal{Q}}\sum_{h=1}^{H}\Bigl[\tau\cdot\max\!\bigl(y_{i,T+h} - \hat{y}_{i,T+h}^{(\tau)},\, 0\bigr) + (1-\tau)\cdot\max\!\bigl(\hat{y}_{i,T+h}^{(\tau)} - y_{i,T+h},\, 0\bigr)\Bigr],
	\label{eq:pinball}
\end{align}
	where $y_{i,T+h}$ is the realized value of series $i$ at horizon $h$. The pinball loss for quantile level $\tau$ penalizes underprediction by a factor $\tau$ and overprediction by a factor $1-\tau$, so minimizing it delivers an estimate of the $\tau$-th conditional quantile of $y_{i,T+h}$ for each series separately, in direct analogy with quantile regression \citep{koenker1978regression}. The median ($\tau = 0.5$) will then serve as the point forecast, while the full set of nine quantiles will be used to construct a predictive distribution which can support interval forecasts and density evaluation. 
    Because the nine quantile levels are estimated jointly but are not constrained to be monotone in $\tau$, the predicted quantiles can in principle cross. We report the model's raw quantile outputs without any post-hoc rearrangement. Since the empirical results in this paper are based solely on the median ($\tau = 0.5$) as the point forecast, quantile crossing does not affect any of the RMSFE comparisons reported below. 
    All series and forecast horizons are produced simultaneously in a single pass through the model. \method uses a hidden state of dimension 128 across its state-weaving stages, yielding approximately one million parameters in total.

\subsection{Model Pretraining}
\label{sec:pretraining}

\citet{ansari2025chronos}  show that a rich mixture of synthetic processes can serve as an effective pretraining source for zero-shot TSFM forecasting. We follow the same idea in the first training stage of \method, which is therefore performed entirely on synthetic data.\footnote{For a broader treatment of synthetic data generation strategies for time-series foundation models, we refer the reader to \cite{ansari2025chronos} and \cite{tempopfn2026}.} We proceed in an iterative way. At each iteration $m = 1, \ldots, M$, where $M$ denotes the total number of synthetic series drawn over the course of pretraining (on the order of ten million), a univariate synthetic series $\{y_t^{(m)}\}_{t=1}^{T_m}$ is drawn from a mixture of generators (more on this below) and split at a randomly chosen cutoff $s_m$. Within this round, the data $\{y_t^{(m)}\}_{t=1}^{s_m}$ is used to estimate the model parameters, and the subsequent observations $\{y_t^{(m)}\}_{t=s_m+1}^{T_m}$ are held out as the evaluation target. The length $T_m$ is itself a random draw from a wide range of values, up to a maximum of 2,048 observations, so that the model is exposed to both short and long histories during training. Next, the model parameters $\boldsymbol{\theta}$ are estimated by minimizing the loss from equation~\eqref{eq:pinball} aggregated across all $M$ synthetic time series. It is worth highlighting here that since every time series is simulated, temporal contamination is ruled out by construction. 

The key idea at this stage is to train the model such that it can acquire broad forecasting skills that do not depend on any particular domain or data-generating process. We build our synthetic data using the generator library introduced by \cite{ansari2024chronos,ansari2025chronos} for the Chronos family of foundation models, but we also expand it in a few directions. The \cite{ansari2025chronos} library relies on three synthetic time series generator families:
\begin{enumerate}
	
	\item \emph{Trend-and-seasonality generators} : each synthetic time series is a multiplicative combination of a trend, a sinusoidal harmonic, and Weibull-distributed noise. The Weibull distribution has support on the positive reals, which is what makes it suitable here: because the noise enters multiplicatively, a positive multiplier scales the trend-and-seasonality template rather than shifting it. The goal of this generator is to produce time series exhibiting patterns that are qualitatively similar to the secular trends and seasonal cycles of economic variables. 
	\item \emph{Sine-wave generators}: each synthetic series consists of a pure sinusoidal oscillation at a configurable period, amplitude, and phase. The goal of this generator is to supply the model with clean periodic templates across a wide range of cycle lengths, exposing it to oscillatory structure in isolation from trend and noise.
	
	\item \emph{Gaussian process and kernel-based generators}: each synthetic time series is drawn from a Gaussian process prior with a covariance kernel that combines squared-exponential, periodic, rational-quadratic, and white-noise components. The goal of this synthetic generator is to cover smooth variation, local roughness, as well as periodicities of different lengths. 
\end{enumerate}

We augment these three with four additional generators adopted from the TempoPFN synthetic-data generation suite~\citep{tempopfn2026}, with the goal of capturing the abrupt changes and mean-reverting dynamics often encountered with macroeconomic data.\footnote{These four generators are used with the default parameterizations of \cite{tempopfn2026}; we summarize the key mechanisms here and refer to their appendix for the full sampling ranges.} 
\begin{enumerate}[resume]
	\item \emph{Sawtooth and step-function generators}: the sawtooth generator produces a periodic ramp, $y_t = A\cdot\mathrm{frac}(t/P + \phi)$ with a randomly chosen ramp direction, period $P$, amplitude $A$, and phase $\phi$, lightly perturbed by a small linear trend and a low-amplitude seasonal term. The step-function generator builds a piecewise-constant series by concatenating subseries of randomized length, each drawn from a configurable menu of local patterns (stable segments, gradual trends, oscillations, or random walks) with its own number of change-points, step sizes, and drift; transitions are optionally Gaussian-smoothed, and global noise, seasonality, and a trend are added on top. The goal of these generators is to expose the model to regime changes and structural breaks of the kind commonly observed in macroeconomic time series.

	\item \emph{Anomaly and spike generators}: these two construct outliers on a baseline series such that the disturbance is the dominant feature. The anomaly generator superimposes spikes that share a common sign within a series, whose magnitudes follow one of several regimes (constant, trending, cyclical, or correlated-random) and whose timing is either evenly spaced, \emph{clustered} (grouped closely together in time), or a mix of the two, with added period jitter. The spike generator places sharp events of a fixed per-series shape (V, inverted-V, or plateau) on a flat baseline in either a clustered ``burst'' mode or an evenly spread mode, with colored noise added probabilistically. The goal of these generators is to train the model to remain robust to sudden, short-lived disturbances such as measurement errors, flash estimates, and one-off shocks that frequently appear in macroeconomic releases.

	\item \emph{CauKer generators}: following the CauKer construction~\citep{xie2025cauker} as adopted in TempoPFN, a structural causal model is sampled over a random directed acyclic graph of $21$ channels (the channel count is inherited from CauKer, not a choice made here). Root nodes are Gaussian-process draws with composite kernels and stochastic mean functions, and each child node applies a nonlinear activation (e.g., ReLU, sigmoid, or sine) to an affine combination of its parents' values, inducing intricate, non-Gaussian dependence. 
    The goal of this generator is to expose the model to the rich, nonlinear temporal patterns that such causal structures generate.
    
	\item \emph{Ornstein--Uhlenbeck generators}: each synthetic series is drawn from an Ornstein--Uhlenbeck process, $dy_t = \theta(t,r_t)\bigl(\mu(t,r_t) - y_t\bigr)\,dt + \sigma(t,r_t)\,dW_t$, simulated by an Euler--Maruyama discretization. The Ornstein--Uhlenbeck process is the continuous-time analogue of a stationary AR(1), so this generator can be read as supplying mean-reverting, AR(1)-like dynamics in continuous time. The latent regime $r_t \in \{0,1\}$ evolves as a two-state Markov chain with fixed but high self-transition probabilities, and \emph{all three} parameters, the mean-reversion speed $\theta$, the long-run mean $\mu$, and the volatility $\sigma$, switch across the two regimes. The goal of this generator is to capture the mean-reverting dynamics and shifting-volatility characteristic of spread and interest-rate series across business-cycle regimes that trend-based or random-walk generators do not reproduce.
\end{enumerate}

As a final step, each synthetic series may pass through a sequence of randomized transformations such as temporal reversal, sign inversion, regime-change insertion, amplitude modulation, seasonal-calendar injection, and resampling artifacts. In addition, some of the synthetic series may be combined into new ones through TSMix-style convex mixing of several sources \citep{darlow2023tsmix}, and missing-value patterns could be imposed. 

Altogether, we generate approximately ten million distinct synthetic forecasting time series, which form the basis of our pretraining. Rather than minimizing equation~\eqref{eq:pinball} over the whole ten million time series at once, the optimization proceeds one step at a time: at each step, a new synthetic series and a new forecast horizon (between one and sixty steps ahead) are drawn from the generator mixture, and the parameters are updated using the pinball loss of equation~\eqref{eq:pinball} evaluated on the held-out horizon. We proceed for 150{,}000 such steps with AdamW, using learning-rate warmup followed by cosine decay. The full run takes about one GPU-day (roughly 28 hours) on an NVIDIA A6000. Importantly, this cost is incurred only once. The resulting weights are then frozen and used as the common initialization for all fine-tuning runs.\looseness=-1

\subsection{Vintage-Consistent Fine-Tuning}
\label{sec:finetuning}

The goal of the synthetic pretraining step is to equip \method with general forecast capabilities that can serve well across a broad range of domains. However, the generality of the synthetic training distributions may not fully replicate the regularities that characterize macroeconomic time series, i.e. factor-like co-movement, persistent autoregressive dynamics, and volatility regimes that shift with the business cycle. For these reasons, we include a fine-tuning step to close this gap, adapting \method to the specific features of macroeconomic data beyond what purely synthetic pretraining can achieve.

It is crucial at this stage that the fine-tuning step does not reintroduce either form of data leakage that the synthetic pretraining was designed to avoid. We address this through vintage-consistent fine-tuning: rather than training on observed macroeconomic data directly, we generate synthetic trajectories from econometric models estimated on real-time data releases, ensuring that neither temporal contamination nor revision bias enters the fine-tuning stage. 
The protection against leakage here comes not from the synthetic nature of the trajectories per se (as was the case for the pretraining stage), but from the fact that the generating models are estimated exclusively on vintage data---the information set actually available to a forecaster at each point in time. The synthetic step serves only to expand an otherwise far-too-small estimation sample.

For this exercise, we rely on the St.~Louis Fed's ALFRED repository, which archives historical snapshots of FRED-MD, each capturing the data as publicly released in a specific month; these snapshots are referred to as vintages. Each vintage $v$ delivers a complete panel $\bm{Y}_{1:T_v}^{(v)} \in \mathbb{R}^{N \times T_v}$, where $N$ is the number of FRED-MD series and $T_v$ the number of months available at that release.\footnote{Our notation reflects the fact that a new vintage does not simply append a row to the previous one. Statistical agencies revise historical observations, which means that the entry $Y_{i,t}^{(v)}$ for series $i$ at date $t$ may differ across vintages $v \neq v'$, and each vintage constitutes an entirely new panel.} At each vintage we apply the FRED-MD transformation codes to obtain the stationary panel $\bm{y}_{1:T_v}^{(v)}$ and interpolate isolated missing observations. This cleaned panel is the sole input to the four generators described below: we estimate the data-generating process from $\bm{y}_{1:T_v}^{(v)}$ and simulate synthetic training trajectories from it, so that revised values and the look-ahead bias they carry with them never enter the training phase, yet the simulated data inherit the macroeconomic structure of the correct historical information set.

We use four generators, each targeting a distinct aspect of macroeconomic data: a dynamic factor model to capture contemporaneous co-movement, Bayesian VARs to capture cross-variable lead--lag effects, univariate AR models to capture series-specific persistence, and a block bootstrap to provide non-parametric coverage of empirical patterns that the parametric models may miss. 

\subsubsection{Dynamic Factor Model Generator}
\label{sec:dfm_gen}

Our first generator builds on a dynamic factor model. There is a large literature documenting how macroeconomic variables share a low-dimensional factor structure, with a small number of latent aggregates  accounting for the bulk of co-movement across series \citep{Stock:Watson:JBES:2002,Giannone:etal:RESTAT:2015}. For each vintage $v$ we estimate the model
\begin{align}
	\bm{y}_t^{(v)} &= \bm{\Lambda}^{(v)}\,\bm{f}_t^{(v)} + \bm{\varepsilon}_t^{(v)}, \label{eq:dfm_obs}\\
	\bm{f}_t^{(v)} &= \bm{A}_1^{(v)}\,\bm{f}_{t-1}^{(v)} + \cdots + \bm{A}_p^{(v)}\,\bm{f}_{t-p}^{(v)} + \bm{u}_t^{(v)},
	\label{eq:dfm_factor}
\end{align}
where $t \in \left\{1,...,T_v\right\}$,   $\bm{f}_t^{(v)} \in \mathbb{R}^k$ is a vector of latent factors, $\bm{\Lambda}^{(v)}$ is the $N \times k$ loading matrix, and the idiosyncratic and factor innovations are mutually independent Gaussians with $\bm{u}_t^{(v)} \sim \mathcal{N}(\bm{0}, \bm{\Sigma}_u^{(v)})$ and $\bm{\varepsilon}_t^{(v)} \sim \mathcal{N}(\bm{0}, \mathrm{diag}(\bm{\sigma}_\varepsilon^{(v)}))$. 
We cast the model in state-space form and estimate the model in equations~\eqref{eq:dfm_obs}--\eqref{eq:dfm_factor} by maximizing the Gaussian likelihood via the expectation-maximization (EM) algorithm, returning the latent factors $\widehat{\bm{f}}_t^{(v)}$ as the Kalman-smoothed states (implemented via \texttt{DynamicFactorMQ}\footnote{\href{https://www.statsmodels.org/stable/generated/statsmodels.tsa.statespace.dynamic_factor_mq.DynamicFactorMQ.html}{https://www.statsmodels.org/stable/generated/statsmodels.tsa.statespace.dynamic\_factor\_mq.DynamicFactorMQ.html}}). We cap the model at $k = 3$ factors, a deliberately small number that, consistent with the macroeconomic factor literature, captures the bulk of the common co-movement across the FRED-MD panel, and fall back to $k = 2$ and then $k = 1$ only on the rare vintages where the richer specification fails to converge. Conditional on the smoothed factors, the loading matrix $\widehat{\bm{\Lambda}}^{(v)}$ is recovered by OLS, and the factor dynamics are estimated by fitting a VAR($p$) with $p = 2$ on $\widehat{\bm{f}}_t^{(v)}$ (an AR($1$) when a single factor is selected).

To generate our synthetic time series, we next draw sequences of factor innovations from $\mathcal{N}(\bm{0}, \widehat{\bm{\Sigma}}_u^{(v)})$, propagate the factors recursively using the fitted VAR coefficients, and recover the observables via the estimated loadings, adding idiosyncratic noise $\bm{\varepsilon}_t^{(s)} \sim \mathcal{N}(\bm{0}, \mathrm{diag}(\widehat{\bm{\sigma}}_\varepsilon^{(v)}))$. We use a 60-step burn-in phase to remove the impact of the initialization, and retain 400 months of synthetic data.

\subsubsection{Bayesian Vector Autoregression Generators}
\label{sec:bvar_gen}

The factor model captures contemporaneous co-movement efficiently, but macroeconomic forecasting also requires capturing the dynamic interdependence through which shocks to one variable drive changes in others over subsequent months. A long-standing literature, going back to {\citet{Doanetal1984} and \citet{Banburaetal2010}, has documented that VARs are well suited to capturing the lead--lag co-movements that are a defining feature of macroeconomic data. To express this structure in our fine-tuning corpus, we estimate Bayesian vector autoregressions (BVARs) on each vintage panel. The specification we rely on is
\begin{align}
	\bm{y}_t^{(v)} = \bm{c}^{(v)} + \bm{B}_1^{(v)}\bm{y}_{t-1}^{(v)} +
	\hdots +
	 \bm{B}_p^{(v)}\bm{y}_{t-p}^{(v)} + \bm{\eta}_t^{(v)},
	\qquad \bm{\eta}_t^{(v)} \sim \mathcal{N}(\bm{0},\,\bm{\Sigma}^{(v)}),
	\label{eq:bvar}
\end{align}
We set $p = 2$ lags and rely on Minnesota-type prior \citep{Litterman1979}. First-lag coefficients are shrunk toward either one (for series where a unit-root prior is appropriate) or zero (for series that appear stationary), while cross-lag coefficients are shrunk toward zero with strength proportional to $\lambda_1\lambda_2\hat\sigma_i/(\hat\sigma_j\ell^d)$, where $\lambda_1 = 0.10$ controls overall tightness, $\lambda_2 = 0.50$ governs cross-lag relative to own-lag shrinkage, and $d = 1$ applies a harmonic decay across lags. 

Because the full FRED-MD panel has $N \approx 126$ variables, estimating a 126-dimensional BVAR on the 200--500 monthly observations available in early vintages calls for careful shrinkage. We therefore use two complementary variants, the first one in particular getting triggered with the earlier vintages: 

\begin{enumerate}
	\item \emph{Clustered BVARs}: we group the $N$ variables using $k$-means clustering, using eight initial groups and splitting any cluster exceeding sixteen variables into sub-groups. We then estimate a separate BVAR for each cluster, using the hyperparameters above. The cluster-specific synthetic time series are then assembled back into a full panel of synthetic data. 
	\item \emph{Full BVAR}: We estimate our BVAR on all 126 variables simultaneously with tighter shrinkage ($\lambda_1 = 0.05$, $\lambda_2 = 0.30$) to regularize the larger parameter space. 
\end{enumerate}
Synthetic panels are simulated from both specifications by recursively applying equation~\eqref{eq:bvar} with multivariate innovations $\bm{\eta}_t^{(s)} = \bm{L}^{(v)}\bm{z}_t$, $\bm{z}_t \sim \mathcal{N}(\bm{0},\bm{I})$, where $\bm{L}^{(v)}$ is the Cholesky of $\widehat{\bm{\Sigma}}^{(v)}$. As with the previous generator, we use a 60-step burn-in phase to remove the impact of the initialization, and retain 400 months of synthetic data. 

\subsubsection{Univariate Autoregressive Generator}
\label{sec:arima_gen}

\citet{Box:Jenkins:1976} established that a large share of the forecastable variation in individual economic series can be captured by low-order univariate autoregressive and moving-average models, a finding that has been repeatedly confirmed in the macroeconomic forecasting literature. To provide \method with synthetic examples of this pattern and to prevent it from only relying on multivariate models, the third generator fits low-order autoregressive models to each transformed series separately. It captures series-specific persistence and innovation scale without imposing common factors or cross-variable lags. 
\begin{align}
	y_{i,t}^{(v)} = \phi_{i,1}^{(v)}\,y_{i,t-1}^{(v)} + \phi_{i,2}^{(v)}\,y_{i,t-2}^{(v)} + e_{i,t}^{(v)},
	\qquad e_{i,t}^{(v)} \sim \mathcal{N}(0,\, \sigma_{i}^{(v)\,2}),
	\label{eq:arima}
\end{align}
where $i = 1, \ldots, N$. The coefficients $(\phi_{i,1}^{(v)}, \phi_{i,2}^{(v)},\sigma_{i}^{(v)\,2})$ are estimated by OLS and synthetic series are simulated recursively as follows
\begin{align}
	y_{i,t}^{(s)} = \sum_{\ell=1}^{p_i}\hat\phi_{i,\ell}^{(v)}\,y_{i,t-\ell}^{(s)} + \hat\sigma_i^{(v)}\,z_{i,t},
	\qquad z_{i,t} \sim \mathcal{N}(0,1),
	\label{eq:arima_sim}
\end{align}
The $N$ series are simulated independently and, after a 60-step burn-in, assembled into a panel. In this way,  the resulting synthetic data have no systematic cross-variable dependence, and the only structure preserved is vintage-calibrated individual own-lag persistence.\footnote{A series falls back to small-variance white noise ($\mathcal{N}(0,\,0.01^2)$) in one of two cases: it has too few observations to fit the autoregression, or a fitted coefficient exceeds $0.99$ to guard against explosive least-squares estimates.
}

\subsubsection{Block-Bootstrap Generator}
\label{sec:bootstrap_gen}

The three synthetic generators above approximate the vintage data-generating process through parametric models. To complement them with non-parametric coverage of empirical patterns that parametric models may miss, we also draw block-bootstrapped panels directly from the vintage data. A single panel of length 400 months is constructed by repeatedly sampling contiguous blocks from $\bm{y}_{1:T_v}^{(v)}$ with random start indices and random block lengths, concatenating until the panel is filled, and then clipping values beyond 25 times the median absolute deviation from the series median to prevent extreme observations from dominating. We use two block-length regimes to cover different co-movement scales: a \emph{standard-block} bootstrap draws block lengths uniformly from $[24, 60]$ months, capturing business-cycle-frequency correlations; a \emph{long-block} bootstrap draws from $[36, 96]$ months, preserving the lower-frequency regime persistence visible across multi-year expansions and contractions. This is a block bootstrap with random block lengths, in the spirit of the stationary bootstrap of \citet{Politis01121994}; note, however, that we draw block lengths uniformly over a fixed range rather than from the geometric distribution of the original stationary bootstrap. Both variants sample exclusively from the same vintage panel as the synthetic generators, so no data released after vintage $v$ enter.

\subsubsection{Fine-Tuning Implementation and Design Rationale}
\label{sec:finetuning_rat}

For each vintage, the fine-tuning corpus is assembled from all four generators: 500 standard block-bootstrap panels, 400 long-block bootstrap panels, 400 ARIMA-style panels, 300 DFM panels, 300 clustered-BVAR panels, and 200 full-BVAR panels, for a total of 2,100 stored panels per vintage. The configuration constants of these generators --- the Minnesota hyperparameters, cluster counts, factor caps, lag orders, burn-in lengths, panel lengths, and the mixture counts above --- are held fixed rather than tuned: they serve to produce plausible, diverse training trajectories for data augmentation, and are deliberately not selected by the statistical criteria one would use for the econometric \emph{benchmarks} of Section~\ref{sec:models}.

The model is not trained on these $2,100$ panels directly. Instead, at each step of the stochastic gradient descent, we build a \emph{new} training example by taking a random slice of one panel and perturbing it slightly. This practice, known in the machine-learning literature as \emph{data augmentation}, is standard when training neural networks. By presenting a slightly different version of the data at every step, it increases the effective number of distinct training examples and discourages the model from memorizing any single panel. Each example is constructed in three steps. Let us denote a stored panel as a matrix $\bm{Y} \in \mathbb{R}^{T \times N}$ with $T = 400$ months and $N = 126$ variables, so $Y_{t,j}$ is the value of variable $j$ in month $t$. First, we draw one of the stored panels, and from that panel we cut a contiguous block of $L = 240$ consecutive months and a random subset $\mathcal{C}$ of $n_c \in \{32, \cdots, 126\}$ variables,
\begin{equation}
  \widetilde{\bm{Y}} = \bm{Y}_{\,t_0:t_0+L,\ \mathcal{C}}, \qquad t_0 \sim \mathrm{Uniform}, \quad \mathcal{C} \subset \{1,\dots,N\},\ \ |\mathcal{C}| = n_c .
  \label{eq:aug_slice}
\end{equation}
The block is then split into a context (history) sub-block and a forecast (horizon) sub-block.
The fine-tuning objective is the same pinball loss of equation~\eqref{eq:pinball} used in pretraining; only the data on which the loss is evaluated changes.

Second, we apply two random perturbations to this slice. We draw a random permutation $\pi$ of the $n_c$ selected variables and reorder the columns accordingly, and we add small Gaussian noise scaled to each variable's own dispersion,
\begin{equation}
    Y^{\mathrm{aug}}_{t,j} = \widetilde{Y}_{t,\pi(j)} + \varepsilon_{t,j}, \qquad \varepsilon_{t,j} \sim \mathcal{N}\!\bigl(0,\ (\kappa\,\mathrm{MAD}_j)^2\bigr), \quad \kappa = 0.05,
    \label{eq:aug_noise}
\end{equation}
where $\mathrm{MAD}_j$ is the median absolute deviation of variable $j$ over the sampled window. The permutation makes the model invariant to the (arbitrary) column ordering of the panel, so it cannot rely on, say, industrial production always occupying a fixed position; the noise, set to $5\%$ of each series' typical deviation, mimics the small measurement errors and data revisions present in real releases and prevents the model from over-fitting to the exact simulated values.

Third, a handful of FRED-MD series (such as the help-wanted indices) are far more volatile. To ensure the model is regularly exposed to these outlier-prone variables, with probability $0.3$ we force them into the sampled subset $\mathcal{C}$.

Fine-tuning runs for 6,000 optimization steps using standard neural network training choices: the AdamW optimizer, a short linear learning-rate warmup, cosine decay, gradient clipping, and mixed-precision arithmetic.\footnote{The optimizer and schedule follow standard practice for fine-tuning large neural networks. \textit{AdamW} \citep{Loshchilov:Hutter:2019} extends Adam \citep{Kingma:Ba:2015} by decoupling weight decay (an L2 penalty that shrinks parameters toward zero to discourage overfitting) from the adaptive gradient scaling, which improves generalization. The \textit{linear warmup} gradually increases the learning rate from zero over the first 250 steps, stabilizing early training before the model has seen much data. \textit{Cosine decay} then smoothly reduces the learning rate to a small floor value over the remaining steps, preventing overshooting near convergence. \textit{Gradient clipping} at $L^2$-norm 1.0 caps the size of each parameter update, protecting against occasional large gradients that would destabilize training. Finally, \textit{bfloat16} is a 16-bit floating-point format that halves memory usage and accelerates computation relative to 32-bit precision, with negligible loss in numerical accuracy for neural network training.} The gradient-based fine-tuning itself completes in roughly nine minutes on a modern GPU. Adding the one time (per vintage) cost of estimating the four generators and simulating their training panels, a complete vintage refresh runs end-to-end in well within forty-five minutes.
This speed is a key feature of the design: because each fine-tuning run completes in minutes, \method can be re-calibrated and re-synchronized at every new data vintage. By contrast, the computational cost of fine-tuning existing TSFMs makes such vintage-by-vintage re-calibration infeasible in practice, forcing practitioners to either use outdated model weights or forgo fine-tuning altogether. The feasibility of re-calibration at each vintage is not a convenience but a prerequisite for genuine real-time forecasting.

\section{Empirical Application}
\label{sec:empirical}

\subsection{Data}
\label{sec:data}

We collect all monthly variables for the United States contained in the
FRED-MD database \citep{McCracken:Ng:2016}, obtained from the Federal Reserve
Bank of St.\ Louis and available at \url{https://fred.stlouisfed.org}.
The database covers a wide range of key macroeconomic variables that applied
economists monitor regularly, including measures of output, labor market
conditions, consumption, housing, money and credit, interest and exchange
rates, prices, and stock market performance.
Of the 137 variables in the database, 14 series are excluded throughout,
leaving 123 series in the evaluation.\footnote{Eight series are excluded
  for being discontinued, not being available every month, or exhibiting extreme measurement irregularities:
  HWI, HWIURATIO, ACOGNO, ANDENOx, NONBORRES, TWEXAFEGSMTHx, UMCSENTx,
  and VIXCLSx.
  A further six series are excluded because their publication lag exceeds
  two months, so their most recent vintage observation lags the forecast
  origin by more than one release cycle and cannot be reliably carried
  forward: CMRMTSPLx, COMPAPFFx, CP3Mx, MZMSL, {S\&P PE ratio},
  and {S\&P div yield}.}

A central feature of our exercise is its real-time character: rather than
using a single, final-vintage dataset, we use \emph{monthly vintages} of
FRED-MD to ensure that each forecast is based exclusively on information
that was publicly available at the time it was made.
These monthly vintages are drawn from the St.~Louis Fed's ALFRED archive, the real-time companion to FRED that stores each historical release (vintage) of the FRED-MD series, as already used for the fine-tuning generators in Section~\ref{sec:finetuning}.\footnote{The composition of the FRED-MD panel has evolved over the sample period: the first vintage (August 1999) contained 112 series, while the most recent vintage in our evaluation window (December 2024) contains 126. Rather than restricting attention to a fixed balanced panel, we expand the forecasting set in step with the official FRED-MD releases: each series enters the evaluation from the first vintage in which it appears, and the same 14 exclusions described above are applied throughout. As a result, the number of series being forecast at any given vintage $t$ reflects exactly the variables available in that release.}
Forecast errors are then measured against the vintage of FRED-MD released
approximately six months after each target date, rather than against the
most recently revised actuals.
This ``six-month delayed actuals'' convention guards against conflating
forecast error with subsequent data revisions while avoiding the noisier
real-time preliminary releases.

\subsection{Models Considered}
\label{sec:models}

We evaluate the following five forecasting models against an AR(1) benchmark:

\begin{itemize}\setlength\itemsep{4pt}

  \item \textbf{AR(1)}: First-order autoregression estimated by OLS at each vintage; this is the benchmark against which all RMSFE ratios are computed.

  \item \textbf{BVAR}: Bayesian VAR with natural conjugate
    Normal--Inverted Wishart prior \citep{KK1993, Kadiyala:Karlsson:1997, Banburaetal2010} and hyperparameters estimated by
    maximizing the marginal likelihood. 

  \item \textbf{DFM}: Factor-augmented direct projections.
    The number of common factors is selected by the Bai--Ng (2002) information
    criterion (up to a maximum of eight), factors are extracted from the balanced
    panel via Principal Components and the EM algorithm \citep{Stock:Watson:JASA:2002},
    and the IC-selected factors are then used to augment a univariate autoregression for each series
    and each horizon \citep{McCracken:Ng:2016}.

  \item \textbf{Chronos-2}: Zero-shot probabilistic forecaster \citep{ansari2025chronos}.
    The model tokenizes numerical time series via quantization into a
    finite vocabulary of bins and pretrains a transformer to predict future
    token distributions autoregressively.

  \item \textbf{Moirai2-Small}: Universal probabilistic foundation model
    \citep{liu2025moirai} pretrained on 27 billion observations spanning nine
    domains.
    The model uses multiple patch-size projection layers and handles
    any-variate inputs natively.

  \item \textbf{\method}: Macro-specific foundation model designed with
    the low-frequency dynamics and heterogeneous persistence typical of
    macroeconomic time series in mind.
\end{itemize}

Chronos-2 and Moirai2-Small are applied in a fully zero-shot manner: the publicly released pretrained weights are used directly, with no macroeconomic fine-tuning and no per-series adaptation. \method{} is likewise zero-shot at the series level --- it is never tuned to any individual target series --- but, as described in Section~\ref{sec:finetuning}, it is re-fine-tuned at each forecast origin on synthetic data drawn from econometric models estimated solely on the data vintage available at that origin. This per-vintage fine-tuning is what makes \method{} vintage-consistent and is the source of its leakage-free guarantee; the two competing TSFMs receive no such adaptation and are evaluated exactly as released.

\subsection{Evaluation Metrics}
\label{sec:metrics}

The evaluation sample runs from August 1999 to December 2024, with the
start date determined by the first monthly vintage for which all
econometric models are available.
At each vintage $t$, we load only the data publicly available as of that
date and apply the FRED-MD transformation codes \citep{McCracken:Ng:2016}
to render each series stationary prior to estimation.\footnote{Two minor
  deviations from a pure real-time protocol are worth noting. First, the
  per-series transformation codes and the list of excluded series
  (Section~\ref{sec:data}) are taken from \citet{McCracken:Ng:2016}, which
  were determined on the full revised sample; fixing them ex ante is
  standard practice but uses information not strictly available at the start
  of the evaluation. Second, the synthetic pretraining stage relies on
  generator families and hyperparameters designed by others
  \citep{ansari2024chronos, tempopfn2026}; because that stage never ingests
  observed macroeconomic series, it introduces no temporal contamination or
  revision bias into \method's forecasts.}

Because FRED-MD exhibits a \emph{ragged edge} --- different series carry
different publication lags, so not all variables are observed through the
same date --- we set the forecast origin to the vintage release month $t$
and train all models on data through $t-1$, using an expanding (recursive)
estimation window that grows as the origin advances.
Series available only through $t-2$ (a two-month publication lag) have
their most recently transformed observation carried forward to fill $t-1$;
series with lags of three or more months are excluded from the analysis
(see Section~\ref{sec:data}).
Each model then produces $h$-step-ahead forecasts for
$h = 0, 1, \ldots, 12$, where $h = 0$ is a \emph{nowcast} for the
vintage month $t$ (whose realized value will not be published until the
following vintage) and $h = 1, \ldots, 12$ are standard multi-step-ahead
forecasts. We proceed recursively until December 2024, generating a time series of
out-of-sample forecast errors for each series, each model, and each horizon.
Forecasts targeting the first and second quarters of 2020 (January through
June 2020) are excluded to prevent the extreme disruption brought about by
the onset of the COVID-19 pandemic from distorting the accuracy
comparison.\footnote{More sophisticated approaches to handling
  pandemic-era observations are discussed in \citet{CCMM:RESTAT:svo} and
  \citet{LenzaPrimiceri:2022:jae:VARafter2020}.}

We evaluate predictive accuracy by computing the ratio of root mean-squared
forecast errors (RMSFE) relative to the AR(1) benchmark:
\begin{equation}
  \mathrm{RMSFE}_{ijh}
  \;=\;
  \sqrt{
    \frac{\displaystyle\sum_{\tau} e^2_{i,j,\tau+h}}
         {\displaystyle\sum_{\tau} e^2_{\mathrm{AR1},j,\tau+h}}
  }, \label{eq:rmsfe}
\end{equation}
where $e_{i,j,\tau+h}$ and $e_{\mathrm{AR1},j,\tau+h}$ denote the forecast
errors of model $i$ and the AR(1) benchmark, respectively, for variable $j$
at origin $\tau$ and horizon $h$, and the sum runs over all evaluation
vintages excluding the COVID period.
Values of $\mathrm{RMSFE}_{ijh}$ below one indicate that model $i$ produces
more accurate point forecasts than the AR(1) for variable $j$ at horizon $h$.

Tables report the \emph{median} RMSFE ratio across series within each
FRED-MD category, which is robust to near-degenerate AR(1) benchmarks in
individual series.
To provide a rough gauge of whether the RMSFE ratios are significantly
different from one, we use the \citet{Diebold:Mariano:1995} $t$-statistic
(DM) with the \citet{Harvey:etal:1997} finite-sample correction.
For each (series, model, horizon) triple, we compute the one-sided
HLN-corrected DM statistic for $H_1$: model $i$ has lower expected squared
loss than AR(1).
The loss differential is $d_t = e^2_{\mathrm{AR1},t} - e^2_{i,t}$; its
long-run variance is estimated using a Bartlett kernel with $h-1$
autocovariance lags.
Our use of the DM test with models that are in some cases nested within the
AR(1) benchmark is a deliberate choice: the Monte Carlo evidence in
\citet{ClarkMcCracken2012, Clark:McCracken:2015} indicates that, for nested
models, the DM test compared against standard normal critical values can be
viewed as a somewhat conservative test of equal accuracy in finite samples,
in the sense of tending to have size modestly below nominal size.
Tables report \emph{majority-vote} stars: *, **, *** indicate that more
than 50\% of series in the category have an HLN-corrected $p$-value below
0.10, 0.05, and 0.01, respectively.
This majority-vote rule is a descriptive summary of how broadly a model
improves on the AR(1) within a category; it is not a joint test of
category-level predictive accuracy and does not control the family-wise
error rate across the series in a group.

\subsection{Results}
\label{sec:results}

Panel~A of Table~\ref{tab:rmsfe_all_6mo} reports median RMSFE ratios relative to AR(1) across all 123 FRED-MD series.
Starting with the econometric benchmarks, the DFM delivers modest improvements over the AR(1) at the shortest horizons (0.977 at $h=0$, 0.979 at $h=1$), with ratios drifting above one beyond $h=3$; the BVAR is essentially at parity (1.007 at $h=0$, 0.991 at $h=1$) and likewise edges above one from $h=3$. Among the \tsfm{}s, \method{} is the best-performing model at every horizon from $h=1$ onward (tied with Chronos-2 at $h=6$), with a median RMSFE ratio of 0.971 at $h=1$ --- the only model to carry majority-vote DM significance ($^*$) at that horizon --- declining gradually to 0.995 at $h=12$.
Chronos-2 is competitive at the very short end ($h=0$: 0.952, $h=1$: 0.977) but converges toward one faster than \method{} as the horizon increases.
Moirai2-Small tracks the AR(1) closely throughout, with ratios near one at all horizons beyond $h=1$.

Panels~B and~C of Table~\ref{tab:rmsfe_all_6mo} report the same exercise for two subsets of macroeconomically interpretable series: a Medium set of 18 headline indicators spanning output, labor, credit, financial, and price conditions (Panel~B), and a Large set of 30 series that adds further depth in inventories, monetary aggregates, and interest rates (Panel~C); full definitions of both sets are provided in Table~\ref{tab:variable_subsets} in~\ref{sec:variable_subsets}.

\begin{table}[!ht]
\centering
\caption{Relative RMSFE vs AR(1) --- all series and curated subsets (6-month delayed actuals)}
\label{tab:rmsfe_all_6mo}
\begin{threeparttable}
\begin{tabular}{lrrrrrr}
\toprule
\multicolumn{1}{l}{} & \multicolumn{6}{c}{Forecast horizon (months ahead)} \\
\cmidrule(lr){2-7}
Model & h = 0 & h = 1 & h = 3 & h = 6 & h = 9 & h = 12 \\
\midrule
\multicolumn{7}{l}{\textit{Panel A: All series (123 series)}} \\
\midrule
BVAR & 1.007 & \textbf{0.991} & 1.017 & 1.013 & 1.019 & 1.029 \\
DFM & \textbf{0.977} & \textbf{0.979} & 1.004 & 1.005 & 1.004 & 1.007 \\
Chronos-2 & \underline{\textbf{0.952}} & \textbf{0.977} & \textbf{0.993} & \underline{\textbf{0.992}} & \textbf{0.998} & \textbf{0.996} \\
Moirai2-Small & \textbf{0.984} & \textbf{0.983} & 1.000 & 1.002 & 1.001 & 1.003 \\
\method & \textbf{0.959}* & \underline{\textbf{0.971}*} & \underline{\textbf{0.987}} & \underline{\textbf{0.992}} & \underline{\textbf{0.995}} & \underline{\textbf{0.995}} \\
\midrule
\multicolumn{7}{l}{\textit{Panel B: Medium variable set (18 series)}} \\
\midrule
BVAR & 1.014 & 1.000 & 1.005 & 1.027 & 1.035 & 1.093 \\
DFM & \textbf{1.000} & \textbf{1.000} & 1.004 & \textbf{1.000} & 1.009 & 1.013 \\
Chronos-2 & \underline{\textbf{0.939}} & \textbf{0.975} & \underline{\textbf{0.965}} & \underline{\textbf{0.967}} & \underline{\textbf{0.993}} & \underline{\textbf{0.991}} \\
Moirai2-Small & \textbf{0.960} & \textbf{0.989} & \textbf{1.000} & \textbf{0.998} & \textbf{1.000} & 1.006 \\
\method & \textbf{0.962}* & \underline{\textbf{0.974}*} & \textbf{0.974} & \textbf{0.980} & \textbf{0.994} & \textbf{0.996} \\
\midrule
\multicolumn{7}{l}{\textit{Panel C: Large variable set (30 series)}} \\
\midrule
BVAR & 1.041 & 1.011 & 1.018 & 1.023 & 1.035 & 1.095 \\
DFM & 1.007 & \textbf{0.999} & 1.008 & \textbf{1.000} & 1.009 & 1.013 \\
Chronos-2 & \underline{\textbf{0.950}*} & \textbf{0.985} & \textbf{0.992} & \underline{\textbf{0.978}} & \textbf{0.999} & \textbf{0.997} \\
Moirai2-Small & \textbf{0.960} & \textbf{0.973} & \textbf{1.000} & \textbf{0.999} & 1.000 & 1.006 \\
\method & \textbf{0.962}* & \underline{\textbf{0.970}*} & \underline{\textbf{0.975}} & \textbf{0.980} & \underline{\textbf{0.998}} & \underline{\textbf{0.996}} \\
\bottomrule
\end{tabular}
\begin{tablenotes}
\footnotesize
\item \textit{Note:} Median RMSFE ratio relative to AR(1). Panel A covers all 123 evaluation series; Panel B the Medium set of 18 headline indicators; Panel C the Large set of 30 series (full definitions in Table~\ref{tab:variable_subsets}). Bold: model beats AR(1). Underline: best (lowest ratio) at that horizon within the panel. Stars based on majority-vote DM test: * (**,***) if $>$50\% of series have HLN-corrected one-sided DM $p<$0.10 (0.05, 0.01) for H$_1$: model beats AR(1). Actuals: 6-month delayed actuals.
\end{tablenotes}
\end{threeparttable}
\end{table}

\FloatBarrier

On the Medium set of 18 headline indicators (Panel~B), \method{} records ratios of 0.974 at $h=1$ (majority-vote significant) and 0.974 at $h=3$, narrowing to 0.996 at $h=12$.
Chronos-2 is the strongest competitor on this set, with 0.965 at $h=3$ and 0.967 at $h=6$, outpacing \method{} over the medium horizon on these 18 series.
Broadening to the Large set of 30 variables (Panel~C), the results revert toward the all-series picture: the sharp improvements seen on the 18-series Medium set fade, and the median ratios for every model move back toward one. \method{} remains the most consistent performer --- lowest ratio at four of the six horizons --- with Chronos-2 marginally ahead only at $h=0$ (0.950 vs 0.962) and $h=6$ (0.978 vs 0.980). As in the full sample, the econometric benchmarks stay at or slightly above the AR(1).

\begin{figure}[t!]
  \centering
  \includegraphics[width=\linewidth]{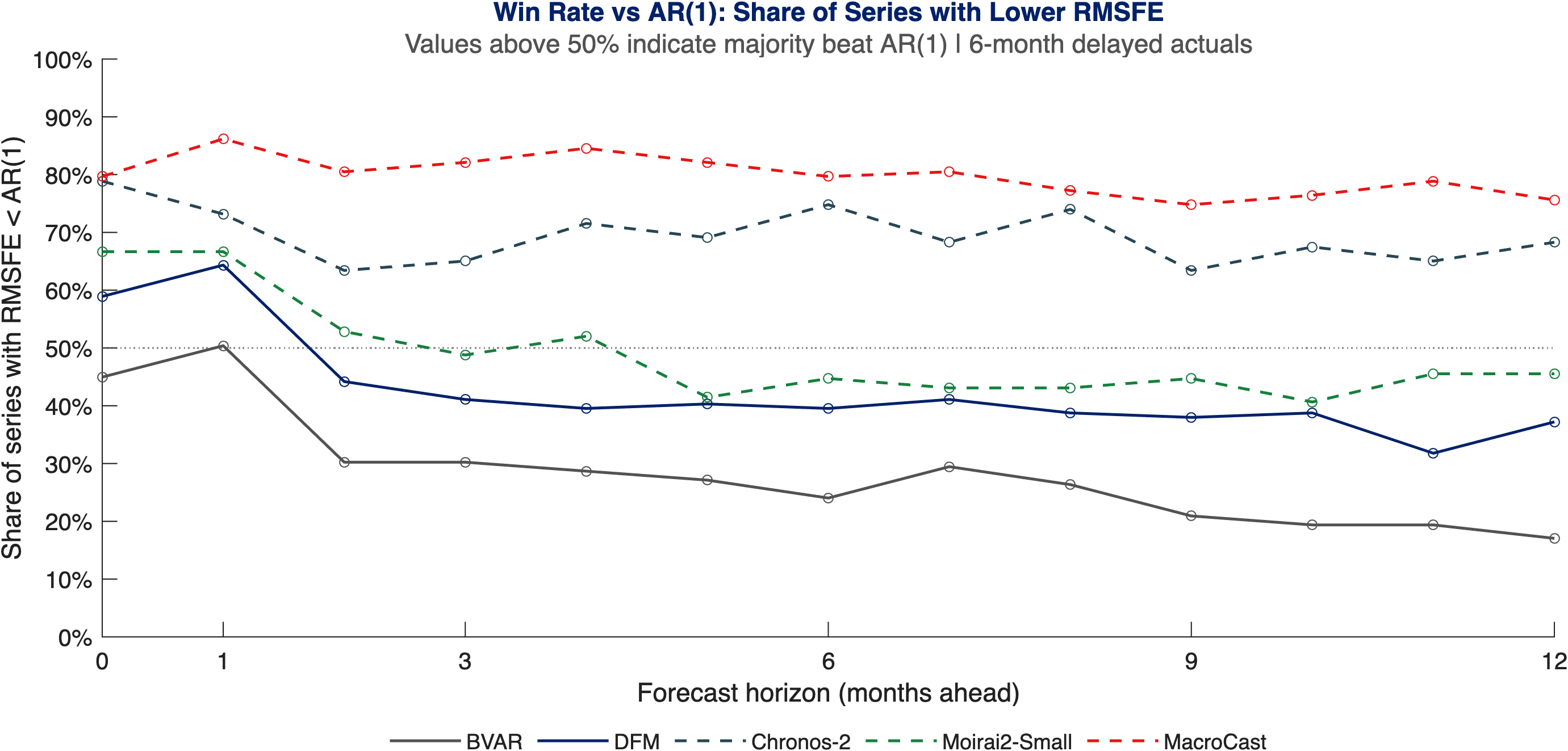}
  \caption{Share of series where each model achieves a lower RMSFE than AR(1),
           by forecast horizon --- \textbf{6-month delayed actuals}.
           Dotted line at 50\%; values above indicate that the majority of
           series beat the benchmark.
           Forecasts targeting Q1--Q2 2020 excluded.}
  \label{fig:win_rate_6mo}
\end{figure}
\FloatBarrier

Figure~\ref{fig:win_rate_6mo} examines the breadth of improvement by plotting the share of 123 series for which each model achieves a lower RMSFE than AR(1).
\method{} beats the benchmark for the majority of series at all horizons from $h=1$ through $h=12$, a pattern not matched by any other model across the full horizon range.
The gap between win rate and statistical significance --- visible for all \tsfm{}s in Figure~\ref{fig:dm_bars_6mo} --- reflects the well-documented difficulty of distinguishing modest but persistent forecast improvements from sampling noise at the individual-series level, particularly over samples of around 300 origins.
Nevertheless, \method{} consistently shows the highest DM significance rates among the \tsfm{}s, and at $h=1$ it is the only model in the group to cross the majority-vote threshold.

\begin{figure}[!t]
  \centering
  \includegraphics[width=\linewidth]{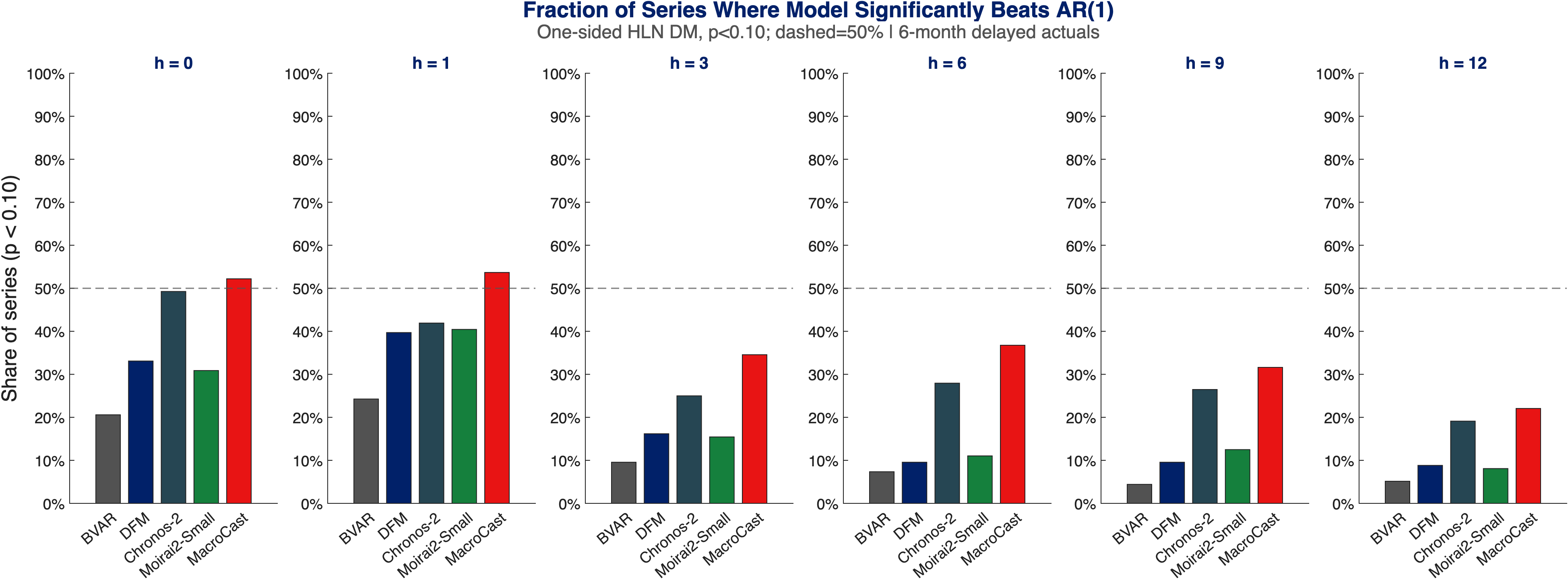}
  \caption{Share of series where the model significantly beats AR(1):
           one-sided HLN-corrected DM test ($p < 0.10$) at
           $h = 1, 3, 6, 9, 12$ --- \textbf{6-month delayed actuals}.
           Dashed line at 50\% (majority-vote threshold used for
           significance stars in the tables).}
  \label{fig:dm_bars_6mo}
\end{figure}

While the median RMSFE ratio and the win rate are informative descriptive statistics, neither offers a full picture: both reduce the cross-section of 123 series to a single summary and discard information about the spread and tails of relative accuracy. Figure~\ref{fig:violin_all_horizons_6mo} addresses this by showing the \emph{entire} cross-sectional distribution of RMSFE ratios, separately for each model and for horizons $h = 1, 3, 6, 12$. Three features stand out.

\begin{figure}[!t]
  \centering
  \includegraphics[width=\linewidth]{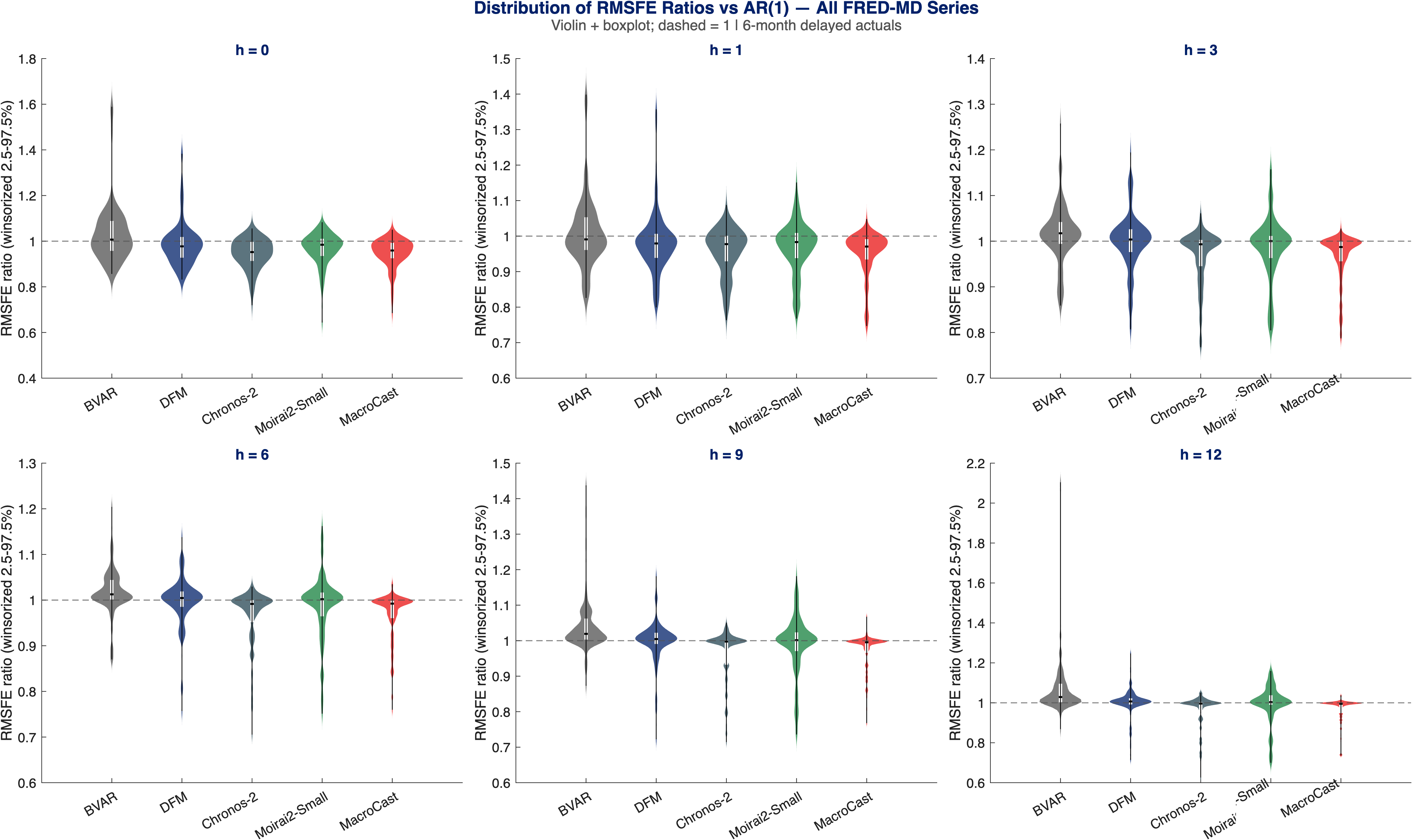}
  \caption{Distribution of RMSFE ratios (relative to AR(1)) across all 123 series at horizons
           $h = 1, 3, 6, 12$ --- \textbf{6-month delayed actuals}.
           Each violin shows the cross-sectional density of series-level RMSFE ratios for one model
           (winsorized at the 2.5th and 97.5th percentiles); the embedded box marks the interquartile
           range and median. Mass below one indicates series for which the model beats the AR(1)
           benchmark. Forecasts targeting Q1--Q2 2020 excluded.}
  \label{fig:violin_all_horizons_6mo}
\end{figure}

First, the models differ sharply in the \emph{spread} of their relative accuracy, not just its center.
\method's distribution is the most tightly concentrated of any model: at $h=1$ roughly 86\% of series fall below the AR(1) benchmark, the bulk of the mass sits just below one, and the upper tail is highly compressed with only about 1\% of series recording a ratio above 1.10.
The econometric benchmarks display the opposite shape, with distributions centered essentially on one and a pronounced \emph{right} tail that lengthens with the horizon, so that by $h=12$ roughly one series in four has a ratio above 1.10 and a handful exceed 1.5. These are the interest-rate and exchange-rate series discussed in Section~\ref{sec:results_timevary}. 

Second, the \tsfm{}s differ in \emph{which} tail is heavy.
Chronos-2 and Moirai2-Small exhibit the longest \emph{left} tails, reaching individual ratios as low as 0.6--0.7 (the housing series), so their occasional gains are the largest of any model; but they also carry more downside dispersion than \method, with a non-trivial share of series above one at the medium horizons.
\method, by contrast, achieves its breadth not through occasional large wins but through a dense concentration of small, reliable improvements coupled with an almost-absent right tail: it is rarely far worse than the AR(1) for \emph{any} series at \emph{any} horizon.

Third, the horizon dimension reveals diverging dynamics.
As $h$ grows the econometric distributions migrate rightward (the BVAR median rises from 0.99 at $h=1$ to 1.03 at $h=12$, and its share of series beating AR(1) falls from 53\% to 17\%), whereas \method's distribution remains anchored below one throughout (its share stays at roughly 75\% even at $h=12$).

\subsection{A Closer Look at Key Macroeconomic Series}
\label{sec:results_timevary}

This section zooms in on five closely watched macroeconomic indicators: industrial production (INDPRO), total nonfarm payroll employment (PAYEMS), capacity utilization in manufacturing (CUMFNS), the unemployment rate (UNRATE), and the federal funds rate (FEDFUNDS). These series span the output, labor market, and monetary policy dimensions of the business cycle, and are among the most scrutinized variables in real-time macroeconomic monitoring.

Table~\ref{tab:rmsfe_tvvars_6mo} reports the RMSFE ratio relative to AR(1) for each model at all six forecast horizons ($h = 0, 1, 3, 6, 9, 12$). Bold entries indicate that the model improves on the AR(1) benchmark for that series and horizon; stars indicate statistical significance from the one-sided HLN-corrected DM test; underlines mark the best-performing model at each horizon.

\clearpage
\begin{table}[!ht]
\centering
\caption{Relative RMSFE vs AR(1) --- Individual target variables (6-month delayed actuals)}
\label{tab:rmsfe_tvvars_6mo}
\begin{threeparttable}
\begin{tabular}{lrrrrrr}
\toprule
\multicolumn{1}{l}{} & \multicolumn{6}{c}{Forecast horizon (months ahead)} \\
\cmidrule(lr){2-7}
Model & h\,=\,0 & h\,=\,1 & h\,=\,3 & h\,=\,6 & h\,=\,9 & h\,=\,12 \\
\midrule
\multicolumn{7}{l}{\textit{Panel A: PAYEMS --- All Employees: Total Nonfarm}} \\
\midrule
BVAR & \textbf{0.964} & \textbf{0.855}$^{***}$ & \textbf{0.925} & \textbf{0.934} & \textbf{0.991} & 1.042 \\
DFM & \textbf{0.977} & \textbf{0.873}$^{***}$ & \textbf{0.913} & \textbf{0.935} & \textbf{0.999} & 1.022 \\
Chronos-2 & \textbf{0.777}$^{**}$ & \textbf{0.849}$^{**}$ & \textbf{0.948} & \textbf{0.933} & \textbf{0.929} & \textbf{0.963} \\
Moirai2-Small & \underline{\textbf{0.714}$^{**}$} & \textbf{0.807}$^{***}$ & \textbf{0.938} & \textbf{0.896} & \textbf{0.951} & 1.009 \\
\method & \textbf{0.731}$^{**}$ & \underline{\textbf{0.747}$^{***}$} & \underline{\textbf{0.851}$^{**}$} & \underline{\textbf{0.849}$^{**}$} & \underline{\textbf{0.859}$^{**}$} & \underline{\textbf{0.900}$^{*}$} \\
\midrule
\multicolumn{7}{l}{\textit{Panel B: INDPRO --- Industrial Production Index}} \\
\midrule
BVAR & \underline{\textbf{0.919}$^{**}$} & \underline{\textbf{0.962}} & 1.028 & 1.054 & 1.043 & 1.040 \\
DFM & \textbf{0.964}$^{*}$ & \textbf{0.967} & 1.016 & 1.018 & 1.011 & 1.023 \\
Chronos-2 & \textbf{0.937}$^{**}$ & 1.011 & 1.009 & 1.015 & 1.015 & 1.017 \\
Moirai2-Small & \textbf{0.959}$^{**}$ & \textbf{0.999} & 1.037 & 1.056 & 1.052 & 1.050 \\
\method & \textbf{0.952}$^{**}$ & \textbf{0.987} & \underline{\textbf{0.974}} & \underline{\textbf{0.992}} & \underline{\textbf{0.998}} & \underline{1.002} \\
\midrule
\multicolumn{7}{l}{\textit{Panel C: UNRATE --- Unemployment Rate}} \\
\midrule
BVAR & \underline{\textbf{0.834}$^{**}$} & \textbf{0.814}$^{**}$ & \underline{\textbf{0.969}$^{**}$} & 1.002 & 1.035 & 1.046 \\
DFM & 1.011 & \textbf{0.938}$^{***}$ & \textbf{0.974} & \textbf{0.994} & 1.013 & 1.001 \\
Chronos-2 & \textbf{0.852}$^{**}$ & \underline{\textbf{0.811}$^{**}$} & \textbf{0.974} & 1.007 & 1.026 & \underline{\textbf{0.991}} \\
Moirai2-Small & \textbf{0.907}$^{*}$ & \textbf{0.817}$^{*}$ & 1.029 & 1.052 & 1.079 & 1.036 \\
\method & \textbf{0.951}$^{**}$ & \textbf{0.958}$^{**}$ & \textbf{0.980}$^{**}$ & \underline{\textbf{0.992}} & \underline{1.001} & 1.001 \\
\midrule
\multicolumn{7}{l}{\textit{Panel D: FEDFUNDS --- Effective Federal Funds Rate}} \\
\midrule
BVAR & 1.649 & 1.557 & 1.258 & 1.098 & 1.080 & 1.126 \\
DFM & 1.375 & 1.576 & 1.130 & 1.040 & 1.038 & 1.038 \\
Chronos-2 & \underline{\textbf{0.847}$^{**}$} & \textbf{0.874}$^{**}$ & \textbf{0.924} & \underline{\textbf{0.939}} & \underline{\textbf{0.980}} & 1.009 \\
Moirai2-Small & \textbf{0.930}$^{**}$ & \textbf{0.896}$^{***}$ & \textbf{0.924}$^{*}$ & \textbf{0.988} & 1.056 & 1.099 \\
\method & \textbf{0.887}$^{***}$ & \underline{\textbf{0.863}$^{***}$} & \underline{\textbf{0.893}$^{***}$} & \textbf{0.952}$^{**}$ & \textbf{0.984} & \underline{\textbf{0.989}} \\
\midrule
\multicolumn{7}{l}{\textit{Panel E: CUMFNS --- Capacity Utilization: Manufacturing}} \\
\midrule
BVAR & \underline{\textbf{0.885}$^{**}$} & \underline{\textbf{0.973}} & 1.030 & 1.042 & 1.031 & 1.018 \\
DFM & \textbf{0.953}$^{*}$ & 1.003 & 1.008 & 1.004 & 1.009 & 1.012 \\
Chronos-2 & \textbf{0.936}$^{**}$ & 1.007 & \textbf{0.996} & 1.001 & 1.001 & 1.004 \\
Moirai2-Small & \textbf{0.955}$^{**}$ & \textbf{0.995} & 1.101 & 1.104 & 1.016 & 1.089 \\
\method & \textbf{0.947}$^{**}$ & \textbf{0.984} & \underline{\textbf{0.975}} & \underline{\textbf{0.992}} & \underline{\textbf{0.994}} & \underline{\textbf{0.996}} \\
\bottomrule
\end{tabular}
\begin{tablenotes}
\footnotesize
\item \textit{Note:} RMSFE ratio relative to AR(1) for individual FRED-MD series. Bold: model beats AR(1). Underline: lowest ratio at that horizon. Stars from HLN-corrected one-sided DM test for the individual series: */(**)/(***) denotes $p<0.10$/$p<0.05$/$p<0.01$ for H$_1$: model beats AR(1). Actuals: 6-month delayed actuals.
\end{tablenotes}
\end{threeparttable}
\end{table}

\clearpage

\paragraph{Nonfarm payroll employment (PAYEMS).}
The most striking results in the table belong to PAYEMS.
\method{} records an RMSFE ratio of 0.747 at $h=1$ --- a 25\% reduction relative to the AR(1) --- and sustains this outperformance throughout the horizon range, reaching 0.849 at $h=6$ and 0.900 at $h=12$, with all estimates statistically significant.
This is the best performance across all models at every horizon from $h=1$ onward by a meaningful margin.
The econometric models are also competitive: BVAR achieves 0.855 and DFM 0.873 at $h=1$, but their advantage fades rapidly after $h=3$.
Moirai2-Small is notably strong at the nowcast and one-step-ahead horizons (0.714 and 0.807) but deteriorates past $h=3$.
The broad and persistent gains for PAYEMS across all models are consistent with the view that employment dynamics carry strong cross-sectional predictability that both multivariate models and \tsfm{}s with sufficiently long context windows can exploit.

\paragraph{Industrial production (INDPRO).}
INDPRO presents a different pattern. The short-run dynamics are well captured by its own lagged values, and the advantages of multivariate models dissipate quickly.
BVAR is competitive at the very short end ($h=0$: 0.919; $h=1$: 0.962) but deteriorates sharply at medium horizons, exceeding one from $h=3$ onward and reaching 1.054 at $h=6$.
The DFM follows a similar, if slightly more muted, trajectory.
Among the \tsfm{}s, only \method{} maintains ratios at or near one throughout the full horizon range (0.987 at $h=1$, 0.974 at $h=3$, 0.992 at $h=6$, 0.998 at $h=9$, 1.002 at $h=12$), making it the only model to track at or below the AR(1) across essentially the full horizon range for this key output series.

\paragraph{Unemployment rate (UNRATE).}
The unemployment rate shows a clear horizon gradient.
At $h=0$--$h=1$, BVAR (0.834; 0.814) and Chronos-2 (0.852; 0.811) are strongly competitive, but both deteriorate past $h=3$, with BVAR and Moirai2-Small recording ratios above one from $h=6$ onward.
\method{} is more conservative at short horizons (0.951; 0.958) but the most stable at medium and long horizons (0.980 at $h=3$, 0.992 at $h=6$), essentially the only model to remain at or below one across the full horizon range (1.001 at $h=9$ and $h=12$).

\paragraph{Federal funds rate (FEDFUNDS).}
BVAR and DFM perform poorly at short horizons, with BVAR recording 1.649 at $h=0$ and 1.557 at $h=1$. This result is only apparently in contrast with earlier findings in the literature, as one has to bear in mind we are applying the BVAR to the series transformed as in \citet{McCracken:Ng:2016}, which enforces stationarity.
By contrast, both \tsfm{}s record strong improvements: Chronos-2 reaches 0.847 at $h=0$ and \method{} 0.863 at $h=1$. \method{} is best at $h=1$, $h=3$, and $h=12$, while Chronos-2 edges ahead at $h=6$ (0.939 vs 0.952) and $h=9$ (0.980 vs 0.984); \method's gains remain statistically significant through $h=6$ (0.952).

\paragraph{Capacity utilization in manufacturing (CUMFNS).}
CUMFNS largely mirrors the INDPRO pattern.
BVAR delivers the best nowcast (0.885) but deteriorates past $h=1$, while \method{} records moderate gains throughout: 0.947 at $h=0$, 0.984 at $h=1$, and best-in-class ratios from $h=3$ (0.975) to $h=12$ (0.996).
The remaining models track the AR(1) closely or exceed it at medium horizons.

\subsubsection{Stability of gains over time.}
The aggregate RMSFE ratios discussed so far average performance over a quarter-century of macroeconomic history that includes the dot-com bust, the Great Recession, a long post-crisis expansion, and the pandemic disruption of 2020. A model that looks competitive in unconditional RMSFE terms may nonetheless exhibit pronounced episodes of under- or over-performance at business cycle turning points, during financial stress, or in the aftermath of structural breaks. To examine the stability of relative performance over time, we track a rolling, 12-month moving-average RMSFE ratio of \method{} relative to AR(1).

For each of the five series we plot this rolling ratio, expressed as \emph{ratio minus one}, separately at each of the six horizons $h = 0, 1, 3, 6, 9, 12$; shaded bands mark NBER recessions. A value below zero (shaded green) indicates that \method{} has lower RMSFE than AR(1) over the preceding twelve months; a value above zero (shaded red) indicates the benchmark is winning over that window. Unlike the full-sample ratios in Table~\ref{tab:rmsfe_tvvars_6mo}, which average over the entire evaluation period, these paths reveal \emph{when} the relative advantage is earned and when it is given back.

Figures~\ref{fig:timevary_indpro}--\ref{fig:timevary_fedfunds} display the results, and three broad patterns emerge. First, relative performance is far from constant: for every series the rolling ratio crosses the zero line repeatedly, confirming that the full-sample averages conceal substantial sign-switching over time.

Second, \method's advantage is concentrated around cyclical turning points and stress episodes. The clearest gains appear for the labor-market and activity series around the 2001 recession and, especially, the 2008--2009 financial crisis, where \method{} opens a wide and sustained lead over AR(1) for PAYEMS, the unemployment rate, and the federal funds rate. During the calm expansions in between --- the mid-2000s and the 2016--2019 stretch --- the parsimonious AR(1) is hard to beat, and the rolling ratio for several series drifts above zero. This pattern is consistent with \method{} capturing nonlinear behavior in periods of turmoil. 

Third, the federal funds rate is the most revealing case. \method{} wins decisively during the 2008--2009 easing, which is what drives its favorable full-sample ratio, but it loses substantially during the 2022--2023 tightening cycle, when the policy rate followed a near-deterministic upward path that the random-walk-like AR(1) benchmark tracks well. The unemployment rate shows a milder version of the same reversal, with \method{} ahead for most of the sample but trailing during the 2022--2024 labor-market cooling. By contrast, the rolling ratios for industrial production and capacity utilization stay close to the zero line throughout, consistent with their near-one full-sample ratios: neither model holds a durable edge for these series. Across all five variables the relative gains are most stable at the shortest horizons ($h = 0, 1$) and become progressively noisier as the horizon lengthens.

\begin{figure}[!t]
  \centering
  \includegraphics[width=0.8\linewidth]{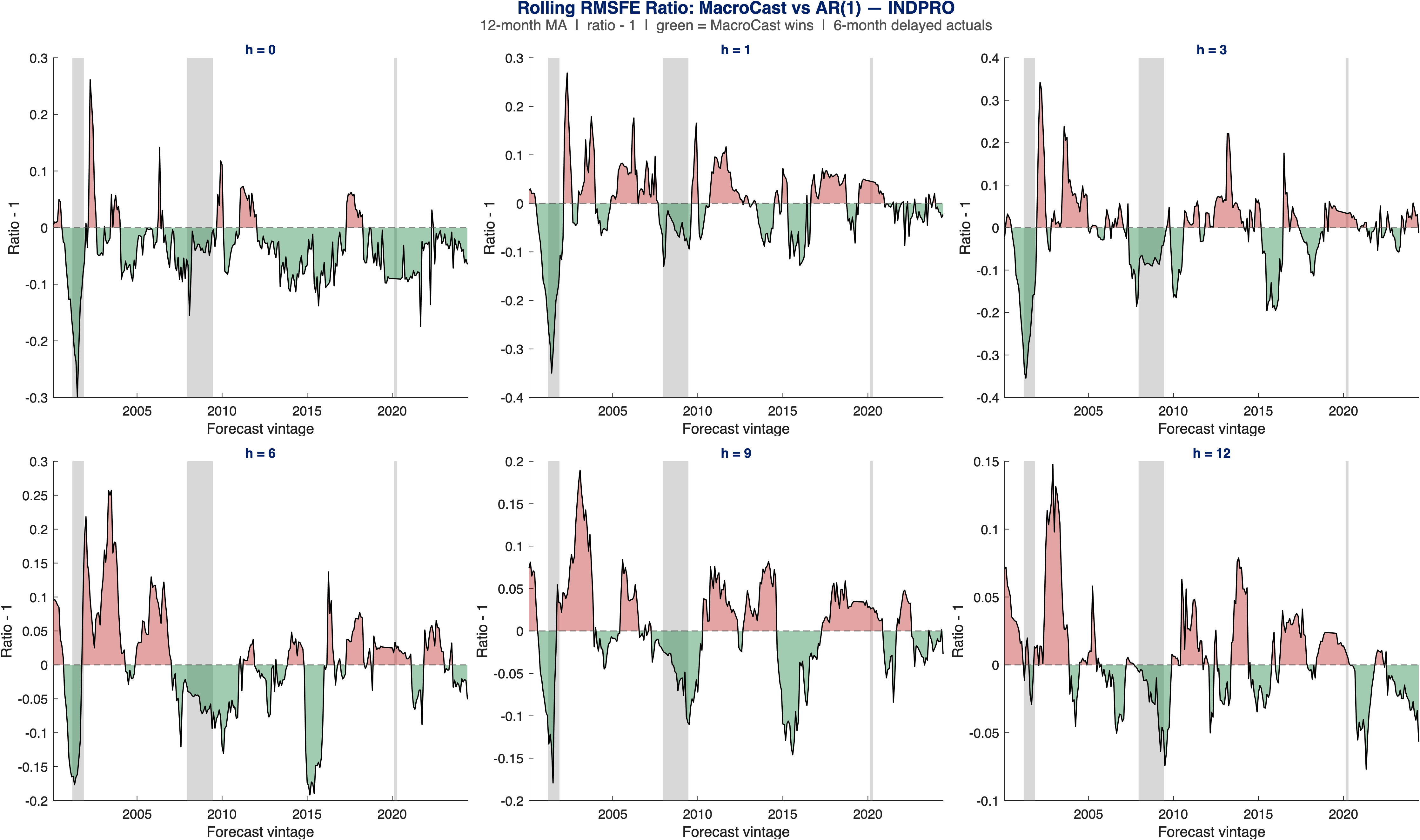}
  \caption{Rolling 12-month RMSFE ratio of \method{} relative to AR(1), shown as ratio minus one ---
           \textbf{Industrial Production (INDPRO)}. Each panel corresponds to a forecast horizon
           $h = 0, 1, 3, 6, 9, 12$. Values below zero (green) indicate that \method{} beats AR(1)
           over the trailing twelve months; values above zero (red) favor the benchmark. Shaded
           bands mark NBER recessions. Forecasts targeting Q1--Q2 2020 excluded.}
  \label{fig:timevary_indpro}
\end{figure}

\begin{figure}[!t]
  \centering
  \includegraphics[width=0.8\linewidth]{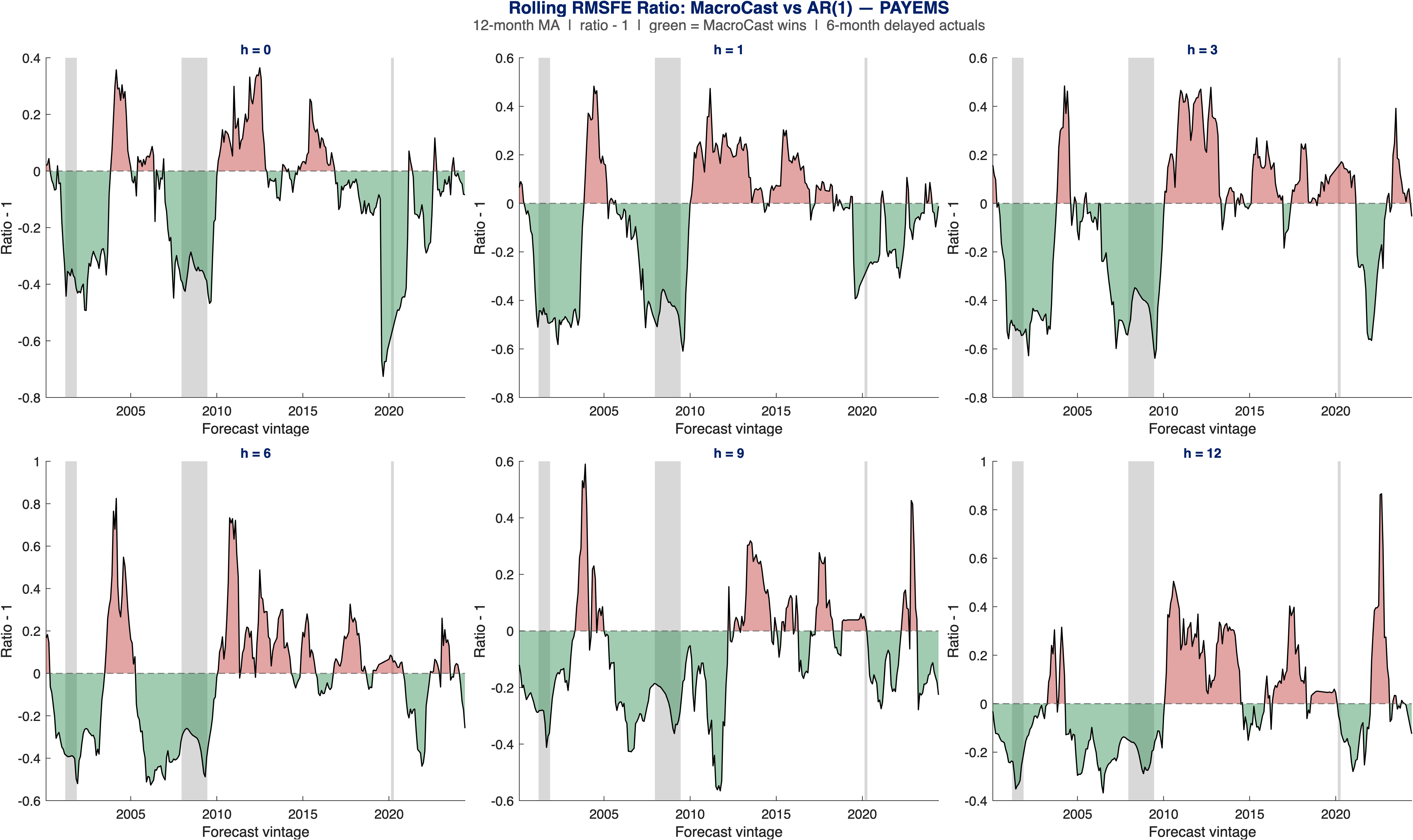}
  \caption{Rolling 12-month RMSFE ratio of \method{} relative to AR(1), shown as ratio minus one,
           by horizon $h = 0, 1, 3, 6, 9, 12$ --- \textbf{Nonfarm Payroll Employment (PAYEMS)}.
           Below zero (green): \method{} beats AR(1). See Figure~\ref{fig:timevary_indpro} for details.}
  \label{fig:timevary_payems}
\end{figure}

\begin{figure}[!t]
  \centering
  \includegraphics[width=0.8\linewidth]{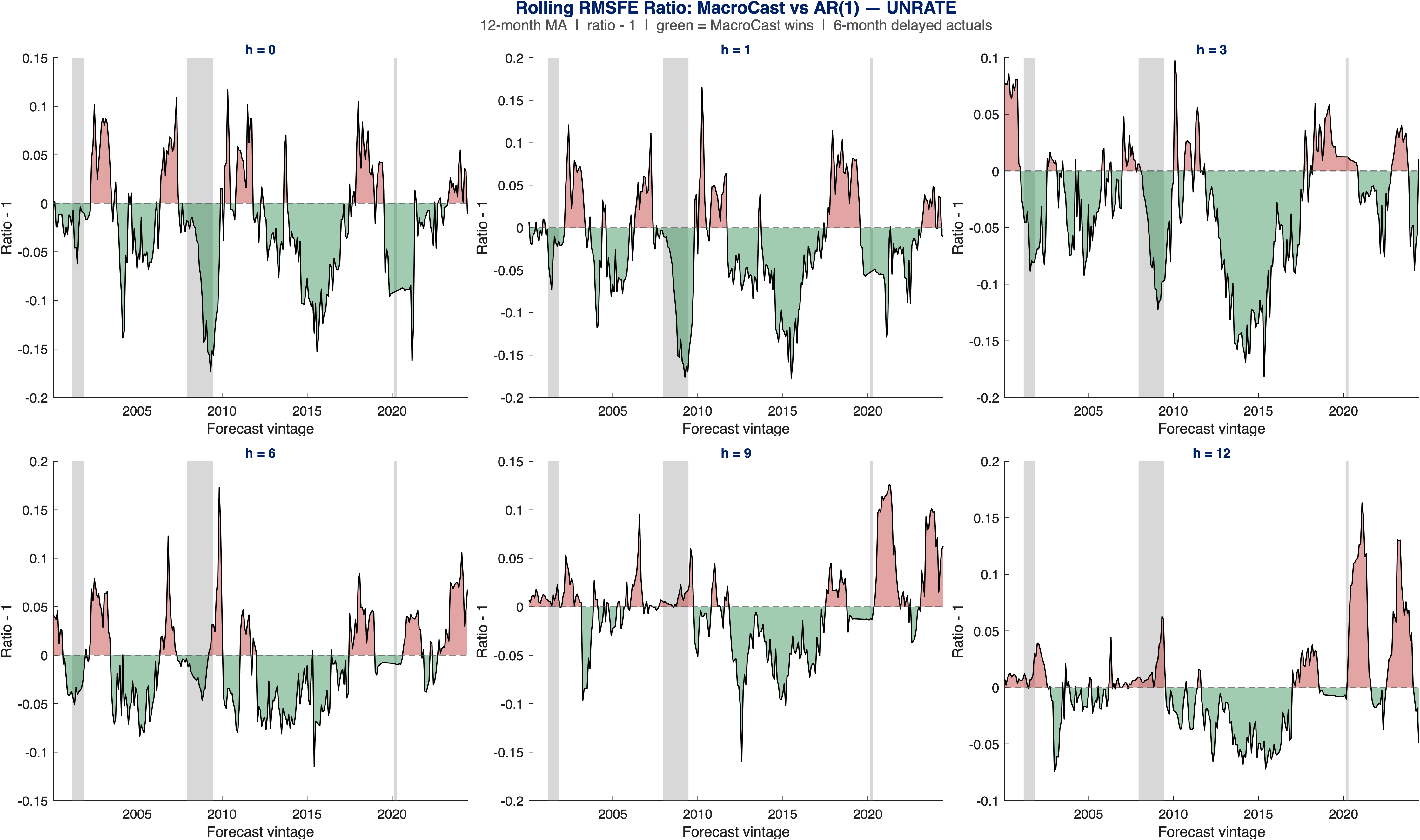}
  \caption{Rolling 12-month RMSFE ratio of \method{} relative to AR(1), shown as ratio minus one,
           by horizon $h = 0, 1, 3, 6, 9, 12$ --- \textbf{Unemployment Rate (UNRATE)}.
           Below zero (green): \method{} beats AR(1). See Figure~\ref{fig:timevary_indpro} for details.}
  \label{fig:timevary_unrate}
\end{figure}

\begin{figure}[!t]
  \centering
  \includegraphics[width=0.8\linewidth]{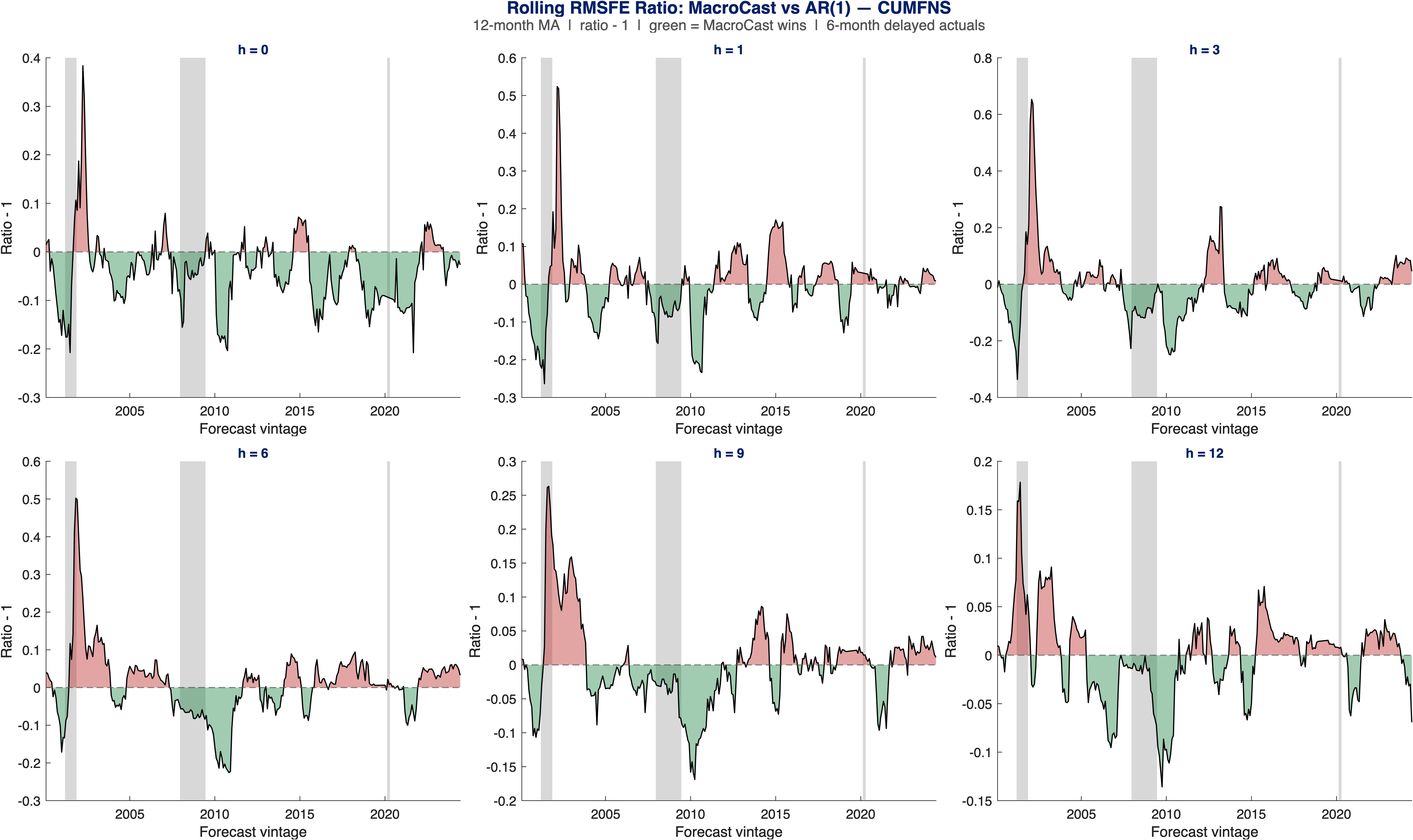}
  \caption{Rolling 12-month RMSFE ratio of \method{} relative to AR(1), shown as ratio minus one,
           by horizon $h = 0, 1, 3, 6, 9, 12$ --- \textbf{Capacity Utilization: Manufacturing (CUMFNS)}.
           Below zero (green): \method{} beats AR(1). See Figure~\ref{fig:timevary_indpro} for details.}
  \label{fig:timevary_cumfns}
\end{figure}

\begin{figure}[!t]
  \centering
  \includegraphics[width=0.8\linewidth]{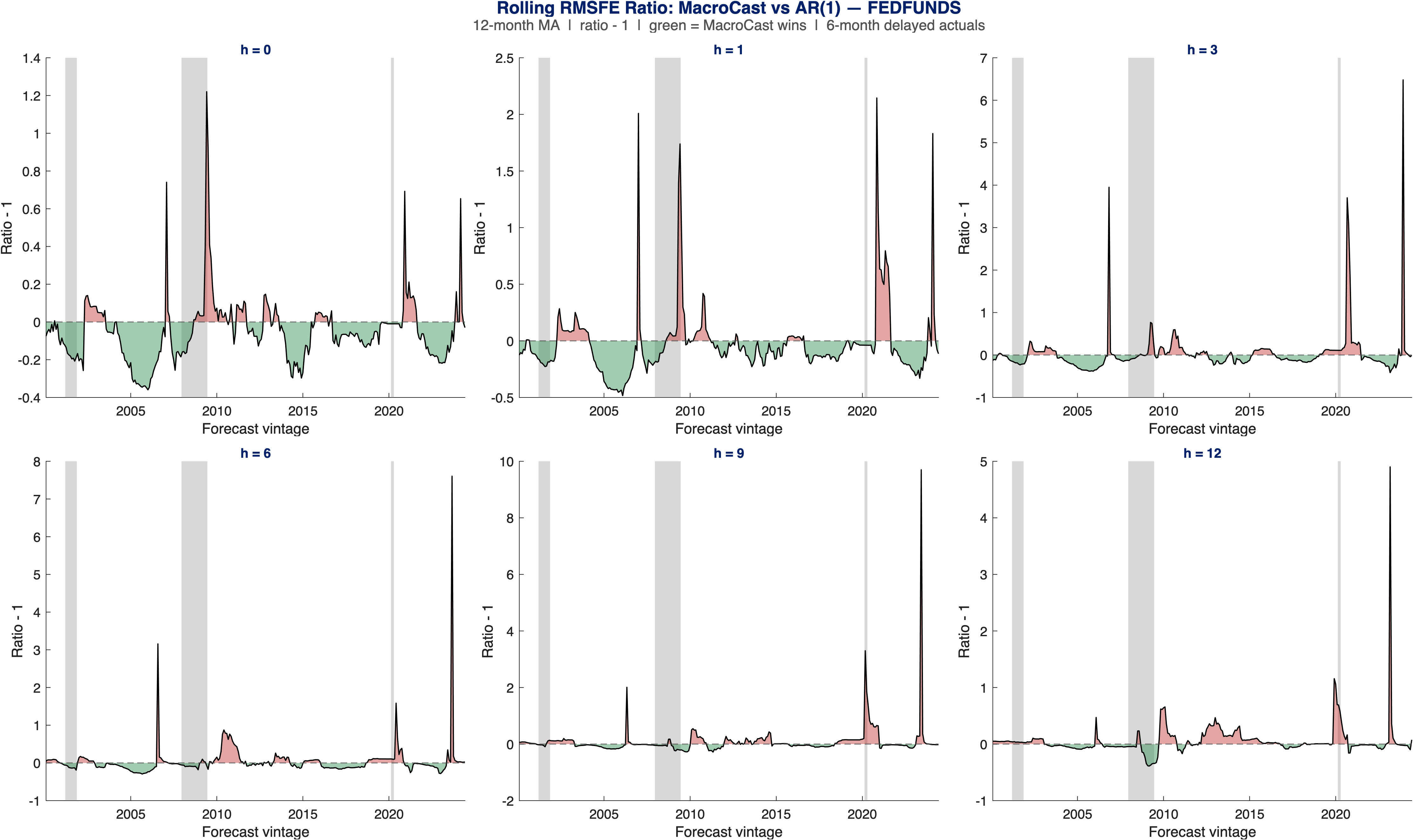}
  \caption{Rolling 12-month RMSFE ratio of \method{} relative to AR(1), shown as ratio minus one,
           by horizon $h = 0, 1, 3, 6, 9, 12$ --- \textbf{Effective Federal Funds Rate (FEDFUNDS)}.
           Below zero (green): \method{} beats AR(1). See Figure~\ref{fig:timevary_indpro} for details.}
  \label{fig:timevary_fedfunds}
\end{figure}
\FloatBarrier
\clearpage

\subsection{Results by FRED-MD Category}
\label{sec:results_bygroup}

This subsection presents the disaggregated results using two complementary views of the same evidence, organized along different dimensions. Table~\ref{tab:rmsfe_bygroup_combined_6mo} is organized by horizon: each panel fixes a forecast horizon ($h = 1, 3, 6, 9, 12$) and reports, for every model, the median RMSFE ratio relative to AR(1) across the series in each of the eight FRED-MD categories. A value below one indicates that the model improves on the AR(1) benchmark for the typical series in that category; stars indicate that more than half of the series in the group pass the one-sided \citet{Harvey:etal:1997} corrected DM test at the conventional significance levels. Table~\ref{tab:rmsfe_bygroup_combined_6mo} is therefore suited to cross-category comparison at a given horizon.

Figures~\ref{fig:dm_bars_bygroup_1_6mo} and~\ref{fig:dm_bars_bygroup_2_6mo} are then organized by category: each of the eight panels fixes one FRED-MD group and reports, for every model, the fraction of series in that group for which the improvement over AR(1) is statistically significant (one-sided HLN-corrected DM test, $p < 0.10$) at each of the five key horizons. These panels are the within-group analog of Figure~\ref{fig:dm_bars_6mo} and are suited to tracking how the statistical reliability of the gains within a given sector evolves as the forecast horizon increases.

Several clear patterns emerge from the disaggregated results, which we discuss in the order in which the categories appear in Figures~\ref{fig:dm_bars_bygroup_1_6mo} and~\ref{fig:dm_bars_bygroup_2_6mo}.

\begin{table}[!ht]
\centering
\caption{Relative RMSFE vs AR(1) by FRED-MD group (6-month delayed actuals)}
\label{tab:rmsfe_bygroup_combined_6mo}
\resizebox{\textwidth}{!}{%
\begin{tabular}{lrrrrrrrr}
\toprule
Model & Output & Labor & Consum. & Housing & Money & Int./FX & Prices & Stocks \\
\midrule
\multicolumn{9}{l}{\textit{Panel A: $h = 1$ month ahead}} \\
\midrule
BVAR & \textbf{0.999} & \textbf{0.976} & \textbf{0.986} & \textbf{0.920} & 1.000 & 1.090 & \textbf{0.972} & 1.070 \\
DFM & \textbf{0.985} & \textbf{0.945} & \underline{\textbf{0.979}} & \textbf{0.936} & \textbf{0.996} & 1.059 & \textbf{0.954} & 1.044 \\
Chronos-2 & 1.002 & \textbf{0.935} & \textbf{0.992} & \underline{\textbf{0.884}} & \textbf{0.978} & 1.001 & \underline{\textbf{0.953}} & 1.035 \\
Moirai2-Small & \textbf{0.995} & \underline{\textbf{0.929}} & \textbf{0.981} & \textbf{0.912} & \textbf{1.000} & \textbf{0.997} & \textbf{0.961} & 1.023 \\
\method & \underline{\textbf{0.984}*} & \textbf{0.958}* & \textbf{0.998}* & \textbf{0.959}* & \underline{\textbf{0.970}*} & \underline{\textbf{0.983}*} & \textbf{0.962}* & \underline{1.000} \\
\midrule
\multicolumn{9}{l}{\textit{Panel B: $h = 3$ months ahead}} \\
\midrule
BVAR & 1.030 & \textbf{0.977} & \textbf{0.998} & \textbf{0.908} & 1.020 & 1.082 & 1.015 & 1.060 \\
DFM & 1.008 & \textbf{0.960} & 1.014 & \textbf{0.962} & 1.003 & 1.096 & 1.003 & 1.025 \\
Chronos-2 & \textbf{0.999} & \underline{\textbf{0.956}} & 1.005 & \underline{\textbf{0.841}} & \textbf{0.993} & \textbf{0.993} & \textbf{0.998} & 1.008 \\
Moirai2-Small & 1.032 & \textbf{0.979} & 1.010 & \textbf{0.853} & \textbf{1.000} & 1.001 & 1.001 & 1.016 \\
\method & \underline{\textbf{0.986}} & \textbf{0.957} & \underline{\textbf{0.998}} & \textbf{0.959} & \underline{\textbf{0.987}} & \underline{\textbf{0.973}} & \underline{\textbf{0.998}} & \underline{\textbf{1.000}} \\
\midrule
\multicolumn{9}{l}{\textit{Panel C: $h = 6$ months ahead}} \\
\midrule
BVAR & 1.056 & \textbf{0.998} & 1.004 & \textbf{0.932} & 1.015 & 1.039 & 1.005 & 1.036 \\
DFM & 1.009 & \textbf{0.982} & 1.002 & \textbf{0.961} & 1.002 & 1.026 & 1.007 & 1.017 \\
Chronos-2 & 1.001 & \textbf{0.981} & \underline{\textbf{0.979}} & \underline{\textbf{0.796}} & \textbf{0.997} & \underline{\textbf{0.982}} & \textbf{0.999} & 1.008 \\
Moirai2-Small & 1.026 & 1.001 & 1.056 & \textbf{0.809} & 1.002 & 1.006 & 1.002 & 1.022 \\
\method & \underline{\textbf{0.992}} & \underline{\textbf{0.955}} & \textbf{0.998} & \textbf{0.967} & \underline{\textbf{0.989}} & \textbf{0.986} & \underline{\textbf{0.998}} & \underline{\textbf{0.995}} \\
\midrule
\multicolumn{9}{l}{\textit{Panel D: $h = 9$ months ahead}} \\
\midrule
BVAR & 1.043 & 1.008 & 1.015 & \textbf{0.981} & 1.026 & 1.042 & 1.008 & 1.039 \\
DFM & 1.009 & \textbf{0.993} & 1.009 & \textbf{0.999} & 1.002 & 1.015 & 1.005 & 1.026 \\
Chronos-2 & 1.001 & \textbf{0.997} & 1.011 & \underline{\textbf{0.772}} & \textbf{0.999} & \underline{\textbf{0.995}} & \textbf{0.998} & 1.011 \\
Moirai2-Small & 1.016 & 1.007 & 1.147 & \textbf{0.785} & \textbf{0.999} & 1.009 & \textbf{1.000} & 1.021 \\
\method & \underline{\textbf{0.995}} & \underline{\textbf{0.966}} & \underline{1.001} & \textbf{0.973} & \underline{\textbf{0.991}} & \textbf{0.996} & \underline{\textbf{0.998}} & \underline{\textbf{0.996}} \\
\midrule
\multicolumn{9}{l}{\textit{Panel E: $h = 12$ months ahead}} \\
\midrule
BVAR & 1.029 & 1.038 & 1.014 & 1.034 & 1.031 & 1.095 & 1.004 & 1.068 \\
DFM & 1.011 & 1.001 & 1.015 & 1.023 & 1.008 & 1.039 & \textbf{0.998} & 1.008 \\
Chronos-2 & 1.008 & \textbf{0.991} & 1.003 & \underline{\textbf{0.762}} & \underline{\textbf{0.992}} & \textbf{0.999} & \textbf{0.997} & 1.006 \\
Moirai2-Small & 1.025 & 1.020 & 1.096 & \textbf{0.792} & \textbf{0.997} & 1.008 & \textbf{0.998} & 1.030 \\
\method & \underline{\textbf{0.998}} & \underline{\textbf{0.971}} & \underline{\textbf{0.999}} & \textbf{0.921} & \textbf{0.994} & \underline{\textbf{0.998}} & \underline{\textbf{0.997}} & \underline{\textbf{0.994}} \\
\bottomrule
\end{tabular}}
\par\smallskip
{\footnotesize
\begin{minipage}{\textwidth}
\textit{Note:} Median RMSFE ratio relative to AR(1) across the series in each FRED-MD group; each panel fixes a forecast horizon. \textbf{Bold} indicates the model beats AR(1); \underline{underline} marks the best (lowest) ratio in each group column within a panel; stars denote majority-vote significance --- more than 50\% of the group's series have an HLN-corrected one-sided DM $p<0.10$ ($^*$), $0.05$ ($^{**}$), $0.01$ ($^{***}$) for H$_1$: model beats AR(1). Evaluation window: August 1999--December 2024, six-month delayed actuals, forecasts targeting Q1--Q2 2020 excluded.
\end{minipage}}
\end{table}

\FloatBarrier

\begin{figure}[tp]
  \centering
  \subfloat[Output \& Income\label{fig:dm_bars_output_income_6mo}]{%
    \includegraphics[width=0.85\linewidth]{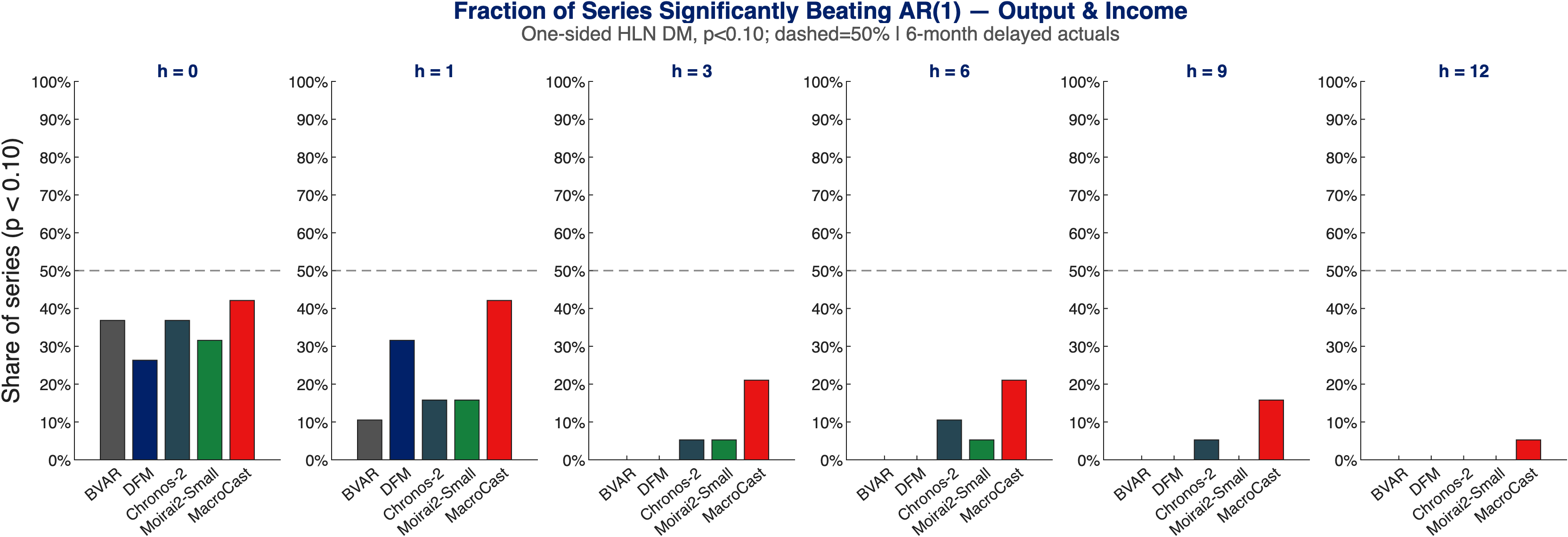}}\\[4pt]
  \subfloat[Labor Market\label{fig:dm_bars_labor_market_6mo}]{%
    \includegraphics[width=0.85\linewidth]{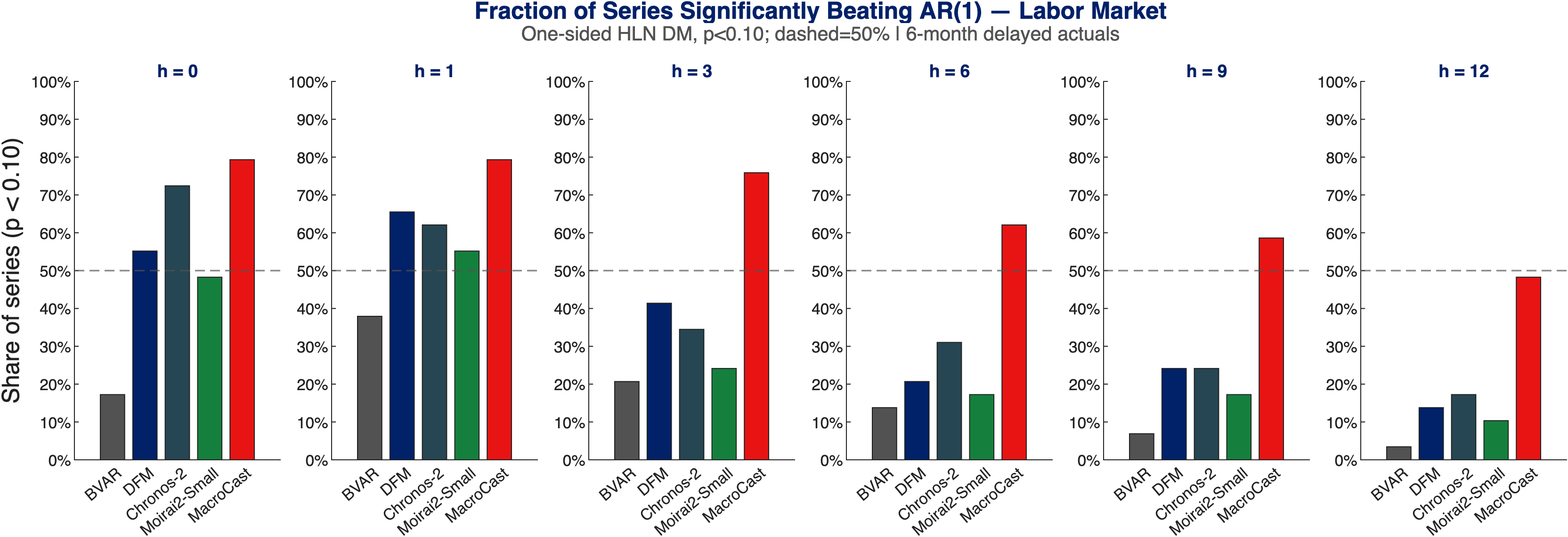}}\\[4pt]
  \subfloat[Consumption \& Inventories\label{fig:dm_bars_consumption_inventories_6mo}]{%
    \includegraphics[width=0.85\linewidth]{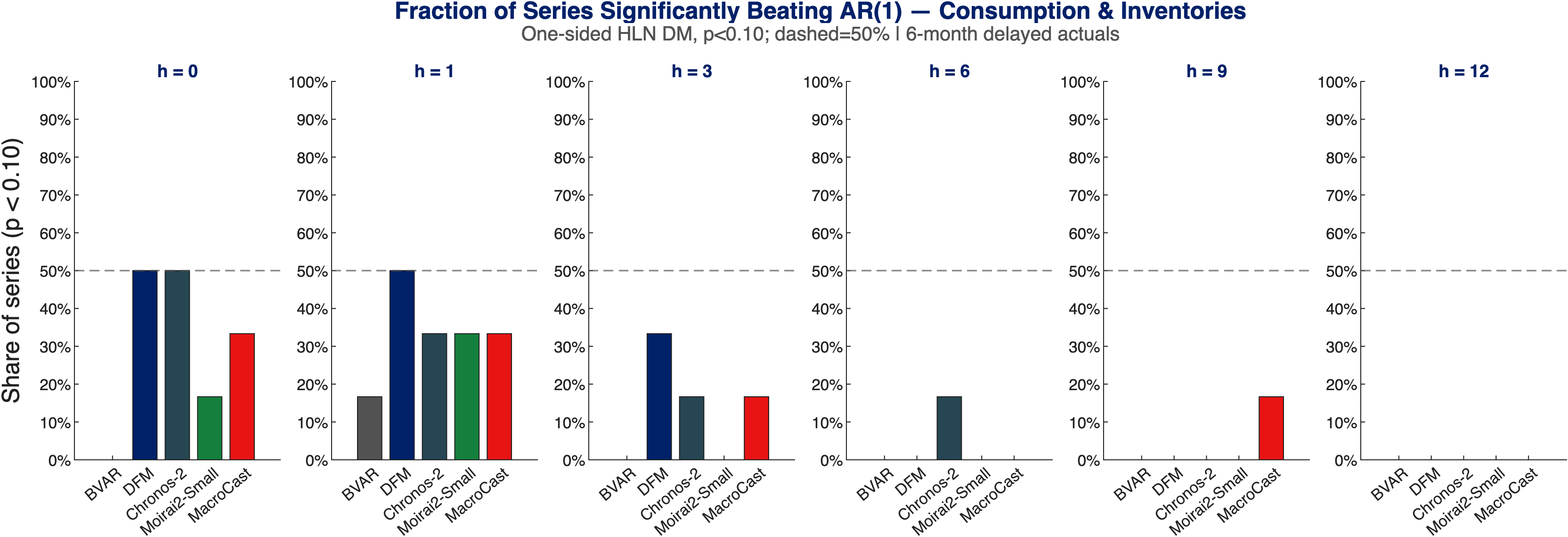}}\\[4pt]
  \subfloat[Housing\label{fig:dm_bars_housing_6mo}]{%
    \includegraphics[width=0.85\linewidth]{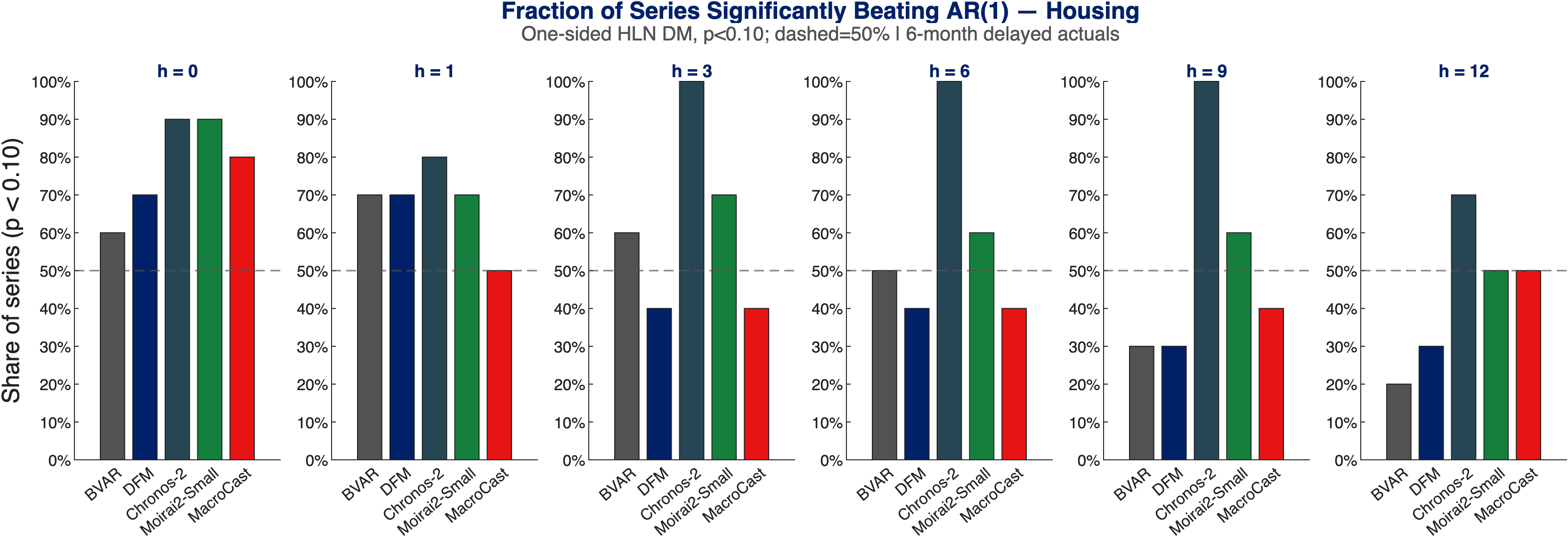}}
  \caption{Fraction of series significantly beating AR(1) by FRED-MD group, panels (a)--(d):
           one-sided HLN-corrected DM test ($p < 0.10$) at $h = 1, 3, 6, 9, 12$ ---
           \textbf{6-month delayed actuals}. Dashed line at 50\%.}
  \label{fig:dm_bars_bygroup_1_6mo}
\end{figure}
\FloatBarrier

\begin{figure}[tp]
  \centering
  \subfloat[Money \& Credit\label{fig:dm_bars_money_credit_6mo}]{%
    \includegraphics[width=0.85\linewidth]{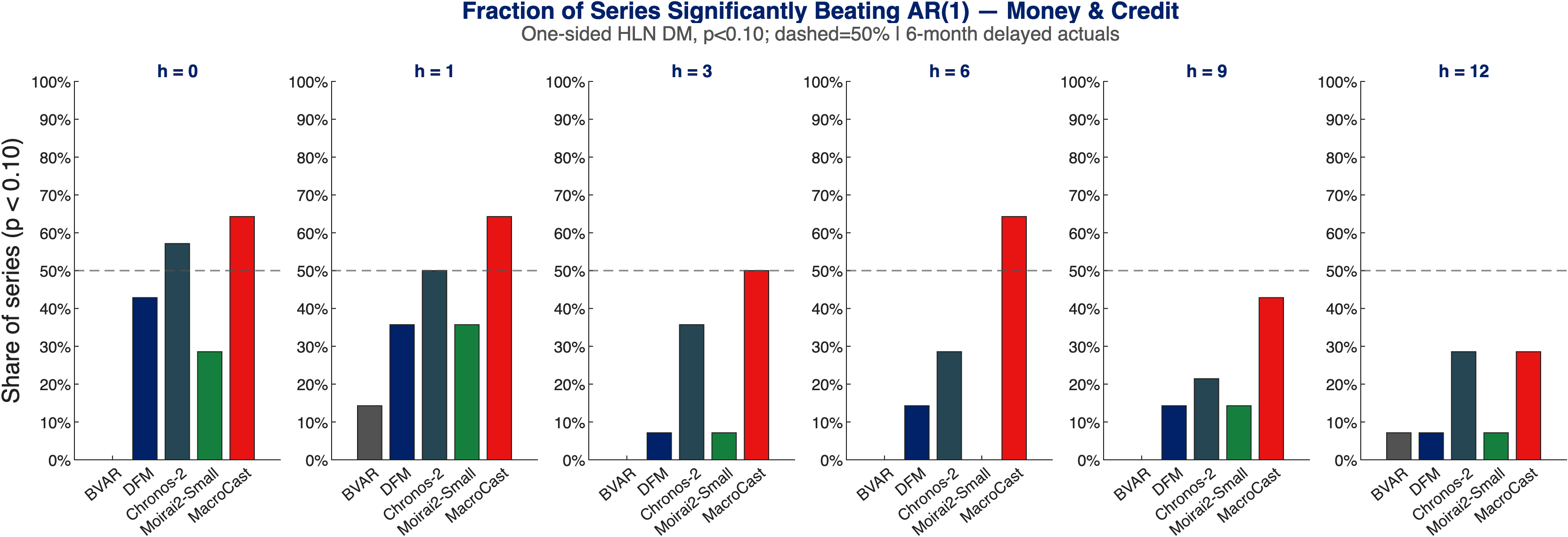}}\\[4pt]
  \subfloat[Interest \& FX Rates\label{fig:dm_bars_interest_fx_rates_6mo}]{%
    \includegraphics[width=0.85\linewidth]{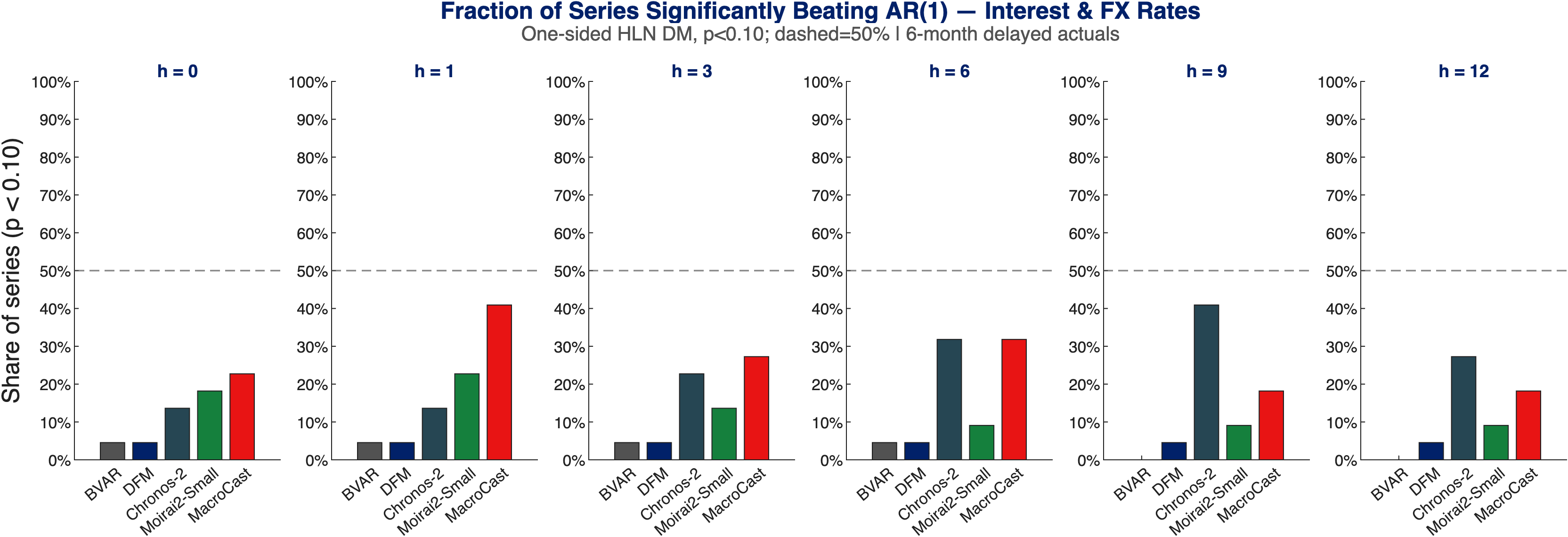}}\\[4pt]
  \subfloat[Prices\label{fig:dm_bars_prices_6mo}]{%
    \includegraphics[width=0.85\linewidth]{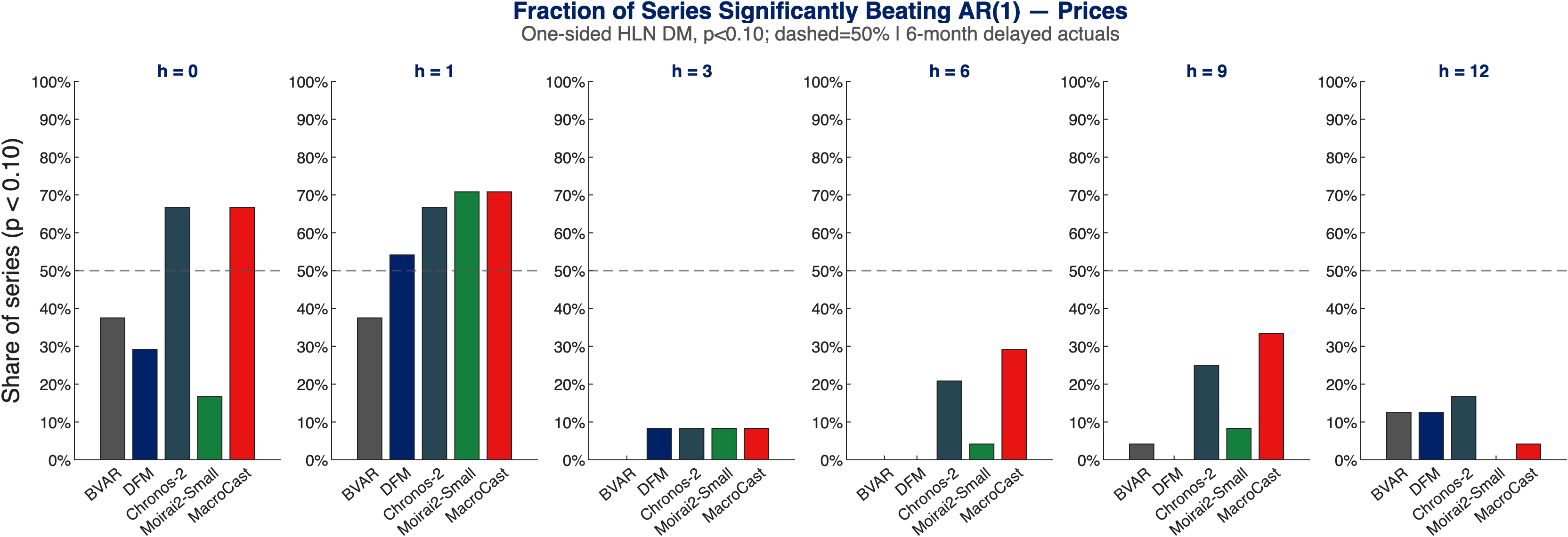}}\\[4pt]
  \subfloat[Stock Market\label{fig:dm_bars_stock_market_6mo}]{%
    \includegraphics[width=0.85\linewidth]{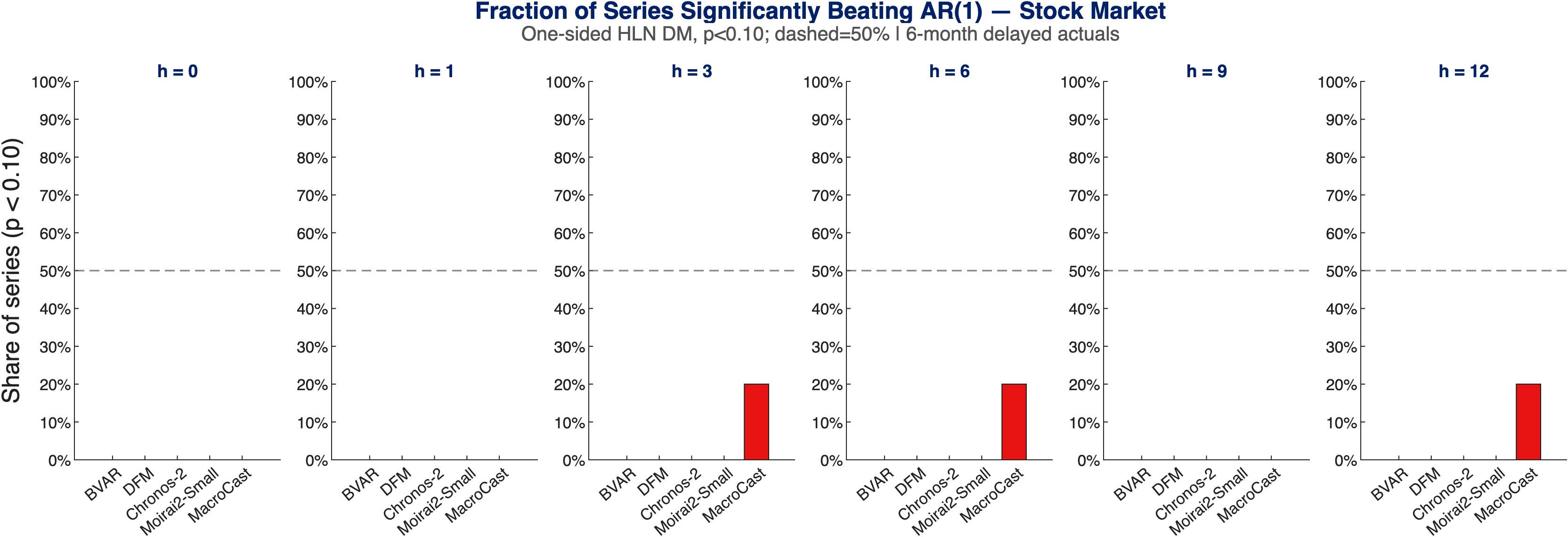}}
  \caption{Fraction of series significantly beating AR(1) by FRED-MD group, panels (a)--(d):
           one-sided HLN-corrected DM test ($p < 0.10$) at $h = 1, 3, 6, 9, 12$ ---
           \textbf{6-month delayed actuals}. Dashed line at 50\%.}
  \label{fig:dm_bars_bygroup_2_6mo}
\end{figure}
\FloatBarrier

\paragraph{Output \& Income.}
For the output and income series (Figure~\ref{fig:dm_bars_bygroup_1_6mo}, panel~(a)), gains over the AR(1) are small but consistent. \method{} is the most reliable model, with best-in-class median ratios from $h=1$ (0.984) through $h=12$ (0.998) and majority-vote significance at $h=1$. BVAR and DFM are competitive only at the shortest horizons and drift above one from $h=3$, consistent with the well-documented robustness of the AR(1) for output growth.

\paragraph{Labor Market.}
Labor market series (Figure~\ref{fig:dm_bars_bygroup_1_6mo}, panel~(b)) are a source of consistent advantage for all models, mirroring the PAYEMS result in Table~\ref{tab:rmsfe_tvvars_6mo}. \method{} and Chronos-2 achieve the strongest overall improvements: \method{} is uniquely best at $h=6$ (0.955), $h=9$ (0.966), and $h=12$ (0.971), while Chronos-2 leads at $h=3$ (0.956). DFM is also competitive across the full horizon range (0.945 to 1.001), performing better here than in most other categories. \method's majority-vote significance at $h=1$ indicates that these gains are statistically reliable for most labor market series.

\paragraph{Consumption \& Inventories.}
For the consumption and inventories group (Figure~\ref{fig:dm_bars_bygroup_1_6mo}, panel~(c)), improvements are again modest. The DFM is strongest at the shortest horizon ($h=1$: 0.979), while \method{} is the most consistent across horizons, with ratios at or just below one throughout and majority-vote significance at $h=1$. Moirai2-Small deteriorates markedly at longer horizons, rising above 1.10 by $h=9$.

\paragraph{Housing.}
Housing (Figure~\ref{fig:dm_bars_bygroup_1_6mo}, panel~(d)) is the category where all models, including the econometric benchmarks, generate the largest and most persistent improvements over the AR(1). Chronos-2 and Moirai2-Small achieve the strongest gains, with Chronos-2 recording median ratios of 0.884, 0.841, 0.796, 0.772, and 0.762 at horizons $h=1$ through $h=12$, a monotonically deepening advantage. Unusually, the gains for housing \emph{strengthen} as the horizon extends rather than fading, suggesting that the univariate AR(1) benchmark becomes progressively weaker at capturing the mean-reverting medium-run dynamics of housing starts and permits. \method{} also improves throughout (0.959 to 0.921), and BVAR and DFM are competitive at short and medium horizons. This is therefore the category where \tsfm{}s most clearly surpass the econometric models in magnitude, though the advantage is concentrated in Chronos-2 and Moirai2-Small.

\paragraph{Money \& Credit.}
For money and credit (Figure~\ref{fig:dm_bars_bygroup_2_6mo}, panel~(a)), \method{} leads at most horizons, with majority-vote significance at $h=1$ (0.970) and best-in-class ratios from $h=1$ to $h=9$; Chronos-2 edges ahead at $h=12$ (0.992 vs 0.994). Gains across all models are modest relative to housing and labor, reflecting the lower cross-sectional predictability of monetary aggregates and credit variables.

\paragraph{Interest \& FX Rates.}
Interest and exchange-rate series (Figure~\ref{fig:dm_bars_bygroup_2_6mo}, panel~(b)) are the category where the econometric models perform most poorly, a direct counterpart of the FEDFUNDS result: BVAR records ratios above 1.08 at short horizons and DFM above 1.05, reflecting the difficulty of fitting near-unit-root series within a cross-sectional shrinkage framework. \method{} is the only model with majority-vote significance ($^*$) at $h=1$ (0.983) and posts the lowest ratio at $h=1$, $h=3$ (0.973), and $h=12$ (0.998), while Chronos-2 edges ahead at $h=6$ and $h=9$.

\paragraph{Prices.}
For the 18 price series (Figure~\ref{fig:dm_bars_bygroup_2_6mo}, panel~(c)), gains over the AR(1) are modest in magnitude but remarkably stable across horizons. \method{} records majority-vote significance at $h=1$ (0.962) and is best from $h=3$ onward, with ratios of around 0.998 throughout. Chronos-2 leads at the very short end ($h=1$: 0.953). BVAR and DFM deteriorate past $h=1$. The pattern is consistent with the near-random-walk behavior of prices at longer horizons, so the improvement opportunities are limited and accrue mostly to models that closely track recent trends.

\paragraph{Stock Market.}
No model beats the AR(1) for stock market variables (Figure~\ref{fig:dm_bars_bygroup_2_6mo}, panel~(d)) with any statistical reliability, and all ratios are close to or above one at every horizon, an unsurprising result given the near-efficiency of asset prices. \method{} is closest to one at every horizon and is the only model to dip below it at $h=6$ and beyond, but the differences are economically negligible.

\paragraph{Overall assessment.}
Taken together, the by-category results sharpen the aggregate picture. \method{}'s outperformance is broad-based: it posts the lowest median ratio at most horizons in Output, Labor, Money and Credit, Interest and FX Rates, and Stock Market, with statistically significant gains in the categories where predictability is highest. The two categories where it trails competitors in raw magnitude (Housing, where Chronos-2 and Moirai2-Small dominate, and Prices at short horizons, where Chronos-2 leads marginally) are also the categories where the leakage-free design exacts the least cost, since both of those advantages could partly reflect the broader corpus of training data in the larger \tsfm{}s. The cross-category evidence therefore reinforces the conclusion that vintage-consistent calibration adds genuine forecasting value, particularly for the series and horizons that matter most for real-time macroeconomic monitoring.

\section{Conclusions}
\label{sec:conclusion}

This paper has introduced \method, a lightweight Time Series Foundation Model designed for real-time macroeconomic forecasting that directly addresses the two forms of data leakage (temporal contamination and revision bias) that undermine the reliability of existing TSFMs for this application. Despite having only one million parameters, one to two orders of magnitude fewer than the foundation models we benchmark against (Moirai2-Small, with roughly 10 million parameters, and Chronos-2, with 120 million) and up to three orders of magnitude fewer than the largest TSFMs in the literature, \method achieves forecast accuracy that is competitive with the best-performing large TSFMs while remaining entirely free of both forms of leakage.

{The approach rests on a two-stage training procedure that navigates a fundamental tension in the TSFM literature. The first stage performs pretraining exclusively on synthetic data. Pretraining is the computationally most intensive part, and using synthetic data only here ensures there is no leakage. Fine-tuning, instead, can be much more lightweight, and at this stage we introduce observational data, crucially in a real-time, vintage-consistent manner. Importantly, this fine-tuning is also grounded in macroeconometric domain knowledge: rather than training on observed series directly, we estimate a suite of econometric models on vintage-specific data, each chosen to reproduce a well-documented feature of the macroeconomy --- dynamic factor models for the low-dimensional factor co-movement across series, Bayesian VARs for cross-variable lead--lag dynamics, and univariate AR specifications for series-specific persistence --- and then train on synthetic draws from those estimated processes.} As a result, \method absorbs the macroeconomic calibration of the correct historical information set without ever ingesting the actual data releases as training targets. The block-bootstrap component supplements this with non-parametric coverage of empirical patterns the parametric models may underfit. The computational cost of this entire procedure is modest: pretraining takes approximately one GPU-day, and each vintage-specific fine-tuning run completes in about nine minutes, making rolling re-estimation over four decades of history entirely feasible.

The empirical results, obtained from a real-time out-of-sample exercise over 123 FRED-MD series and roughly 300 monthly forecast origins spanning August 1999 to December 2024, paint a consistent picture. \method matches or surpasses the best-performing large TSFMs across horizons and variable groups, while also outperforming traditional econometric benchmarks including the BVAR and the factor model.

While our approach eliminates both forms of leakage identified in the paper, a natural extension is to combine \method with real-time nowcasting data or mixed-frequency information, broadening its applicability to settings where the timing of data releases is itself informative.

\bibliographystyle{chicago}
\bibliography{All_references_LLMs}

\clearpage
\begin{appendices}
	\renewcommand{\thesection}{Appendix \Alph{section}}
	\renewcommand{\thesubsection}{\Alph{section}.\arabic{subsection}}
	\renewcommand{\theequation}{\Alph{section}.\arabic{equation}}
	\renewcommand\thetable{\Alph{section}.\arabic{table}}
	\renewcommand\thefigure{\Alph{section}.\arabic{figure}}
	\setcounter{equation}{0}
	\setcounter{section}{0}
	\setcounter{table}{0}
	\setcounter{figure}{0}

    \section{Econometric Model Specifications}
\label{sec:model_specs}

\subsection{Bayesian VAR}
\label{sec:bvar}

We consider Bayesian VARs with natural conjugate Normal--Inverted Wishart
(N-IW) priors.
This prior dates back to \citet{Zellner1971} and was later studied by
\citet{KK1993, Kadiyala:Karlsson:1997}.
\citet{Banburaetal2010} show that it can be successfully
applied to a very large cross-section of macroeconomic data, and several
contributions have followed using this model and prior to handle large
macroeconomic datasets.

Collect $N$ different variables in the vector $y_t = (y_{1t},\ldots,y_{Nt})'$
and write the Vector Autoregression of order $p$, $\mathrm{VAR}(p)$, as
\begin{equation}
  y_t = \Phi_c + \Phi_1 y_{t-1} + \Phi_2 y_{t-2} + \cdots + \Phi_p y_{t-p}
        + \varepsilon_t, \qquad
  \varepsilon_t \sim \mathrm{i.i.d.}\;\mathcal{N}(0,\Sigma). \label{eq:var}
\end{equation}
With $p$ lags and $N$ variables, each equation has $K = Np + 1$ regressors.
Grouping the coefficient matrices in the $N \times K$ matrix
$\Phi' = [\Phi_c\;\Phi_1\;\cdots\;\Phi_p]$ and defining
$x_t = (1, y'_{t-1}, \ldots, y'_{t-p})'$, the $\mathrm{VAR}(p)$ can be
written in compact matrix form as
\begin{equation}
  Y = X\Phi + E, \label{eq:varmat}
\end{equation}
where $Y = [y_1,\ldots,y_T]'$, $X = [x_1,\ldots,x_T]'$, and
$E = [\varepsilon_1,\ldots,\varepsilon_T]'$ are $T\times N$, $T\times K$,
and $T\times N$ matrices, respectively.

The conjugate N-IW prior elicits the prior on the coefficients conditionally
on the error variance $\Sigma$:
\begin{equation}
  \Phi \mid \Sigma \sim \mathcal{N}(\Phi_0,\,\Sigma\otimes\Omega_0),
  \qquad
  \Sigma \sim \mathcal{IW}(S_0, v_0). \label{eq:prior}
\end{equation}
The conditional posterior distribution is also N-IW:
\begin{equation}
  \Phi \mid \Sigma, Y \sim \mathcal{N}(\bar\Phi,\,\Sigma\otimes\bar\Omega),
  \qquad
  \Sigma \mid Y \sim \mathcal{IW}(\bar S, \bar v), \label{eq:posterior}
\end{equation}
where $\bar\Phi = \bar\Omega(\Omega_0^{-1}\Phi_0 + X'Y)$,
$\bar\Omega = (\Omega_0^{-1} + X'X)^{-1}$,
$\bar S = S_0 + Y'Y + \Phi_0'\Omega_0^{-1}\Phi_0 - \bar\Phi'\bar\Omega^{-1}\bar\Phi$,
and $\bar v = v_0 + T$.

We elicit $\Phi_0, \Omega_0, S_0, v_0$ in the Minnesota tradition.
To these priors we add the ``sum of coefficients'' and ``single unit root''
priors of \citet{Doanetal1984} and \citet{Sims:1993:nber:chapter}.
All hyperparameters are estimated by maximizing the marginal likelihood of
the model, which is available in closed form.
In the empirical exercise we set $p = 6$ lags and estimate the BVAR on the
full cross-section of series available at each vintage.
Point forecasts are obtained by taking the mean of the predictive density;
for $h > 1$, the predictive density is produced by iterating the model
forward via Monte Carlo simulation with 5{,}000 draws.

\subsection{Factor Model}
\label{sec:dfm}

Factor models have repeatedly been shown to be well suited for macroeconomic
forecasting \citep{Stock:Watson:2006, McCracken:Ng:2016}.
We follow the implementation of \citet{McCracken:Ng:2016}.
First, the number of common factors $\hat k$ is selected by the information
criterion IC$_{p2}$ of \citet{Stock:Watson:JASA:2002}, with the maximum
number of factors set to eight.
The $\hat k$ factors are then extracted from the balanced panel using
Principal Components Analysis, with the EM algorithm of
\citet{Stock:Watson:JASA:2002} used to fill in any missing values before
extraction.
Second, the IC-selected factors are used to augment a direct-projection
autoregression for each series $i$:
\begin{equation}
  y_{i,t} = \alpha^h + \bm{\beta}^h(L)'\,\hat{\bm{f}}_{t-h}
             + \gamma^h(L)\,y_{i,t-h} + \varepsilon_{i,t}, \label{eq:dfm}
\end{equation}
where $\hat{\bm{f}}_{t}$ is the $\hat k \times 1$ vector of estimated common factors
and $\bm{\beta}^h(L)$ is the corresponding lag polynomial.
This model is estimated via OLS, with the number of lags of $\hat{\bm{f}}_{t}$
and $y_{i,t}$ selected by BIC.
The $h$-step-ahead forecast is then computed as
\begin{equation}
  \hat y_{i,t+h} = \hat\alpha^h + \hat{\bm{\beta}}^h(L)'\,\hat{\bm{f}}_{t}
                   + \hat\gamma^h(L)\,y_{i,t}. \label{eq:dfmfcst}
\end{equation}
This \emph{direct projection} approach minimizes the $h$-step-ahead forecast
error directly, and requires a separate regression for each forecast horizon.

    \section{Variable Subsets}
    \label{sec:variable_subsets}

    Table~\ref{tab:varlist_all} lists all FRED-MD variables considered in this study, organized by the eight categories of \citet{McCracken:Ng:2016}. Series marked with a dagger (${}^{\dagger}$) are excluded from the evaluation dataset (see Section~\ref{sec:data}). For each series we report the mnemonic used in the FRED-MD files and the transformation code applied before model estimation.

        Table~\ref{tab:variable_subsets} lists the series in the Medium (18 series) and Large (30 series) subsets used in Panels~B and~C of Table~\ref{tab:rmsfe_all_6mo}. Variable codes and transformations follow \citet{McCracken:Ng:2016}. The Large set is the union of the Medium set and the 12 additional series in Panel~B.

    {\small
\begin{longtable}{p{4.0cm}p{9.0cm}c}
\caption{FRED-MD variables: mnemonics, descriptions, and transformations}\label{tab:varlist_all} \\
\toprule
Mnemonic & Description & Trans.\\
\midrule
\endfirsthead
\multicolumn{3}{l}{\footnotesize\textit{(continued from previous page)}} \\
\toprule
Mnemonic & Description & Trans.\\
\midrule
\endhead
\midrule
\multicolumn{3}{r}{\footnotesize\textit{(continued on next page)}} \\
\endfoot
\bottomrule
\multicolumn{3}{p{12.5cm}}{\footnotesize\textit{Note:} Trans.\ reports the FRED-MD transformation code: 1~=~level; 2~=~$\Delta$; 3~=~$\Delta^2$; 4~=~$\ln$; 5~=~$\Delta\ln$; 6~=~$\Delta^2\ln$. Series marked ${}^{\dagger}$ are excluded from the evaluation dataset (see Section~\ref{sec:data}).} \\
\endlastfoot
\multicolumn{3}{l}{\textit{Output \& Income}} \\
\midrule
\texttt{INDPRO} & IP Index & $\Delta\ln$ \\
\texttt{IPFINAL} & IP: Final Products (Market Group) & $\Delta\ln$ \\
\texttt{IPCONGD} & IP: Consumer Goods & $\Delta\ln$ \\
\texttt{IPDCONGD} & IP: Durable Consumer Goods & $\Delta\ln$ \\
\texttt{IPNMAT} & IP: Nondurable Materials & $\Delta\ln$ \\
\texttt{IPDMAT} & IP: Durable Materials & $\Delta\ln$ \\
\texttt{IPMAT} & IP: Materials & $\Delta\ln$ \\
\texttt{IPMANSICS} & IP: Manufacturing (SIC) & $\Delta\ln$ \\
\texttt{IPFPNSS} & IP: Final Products and Nonindustrial Supplies & $\Delta\ln$ \\
\texttt{IPNCONGD} & IP: Nondurable Consumer Goods & $\Delta\ln$ \\
\texttt{IPBUSEQ} & IP: Business Equipment & $\Delta\ln$ \\
\texttt{IPB51222S} & IP: Residential Utilities & $\Delta\ln$ \\
\texttt{IPFUELS} & IP: Fuels & $\Delta\ln$ \\
\texttt{CUMFNS} & Capacity Utilization:  Manufacturing & $\Delta$ \\
\texttt{W875RX1} & Real personal income ex transfer receipts & $\Delta\ln$ \\
\texttt{DPCERA3M086SBEA} & Real personal consumption expenditures & $\Delta\ln$ \\
\texttt{DDURRG3M086SBEA} & Personal Cons.  Exp:  Durable goods & $\Delta^2\ln$ \\
\texttt{DNDGRG3M086SBEA} & Personal Cons.  Exp:  Nondurable goods & $\Delta^2\ln$ \\
\texttt{DSERRG3M086SBEA} & Personal Cons.  Exp:  Services & $\Delta^2\ln$ \\
\addlinespace[4pt]
\multicolumn{3}{l}{\textit{Labor Market}} \\
\midrule
\texttt{PAYEMS} & All Employees:  Total nonfarm & $\Delta\ln$ \\
\texttt{CE16OV} & Civilian Employment & $\Delta\ln$ \\
\texttt{USGOOD} & All Employees:  Goods-Producing Industries & $\Delta\ln$ \\
\texttt{USCONS} & All Employees:  Construction & $\Delta\ln$ \\
\texttt{MANEMP} & All Employees:  Manufacturing & $\Delta\ln$ \\
\texttt{DMANEMP} & All Employees:  Durable goods & $\Delta\ln$ \\
\texttt{NDMANEMP} & All Employees:  Nondurable goods & $\Delta\ln$ \\
\texttt{SRVPRD} & All Employees:  Service-Providing Industries & $\Delta\ln$ \\
\texttt{USFIRE} & All Employees:  Financial Activities & $\Delta\ln$ \\
\texttt{USTRADE} & All Employees:  Retail Trade & $\Delta\ln$ \\
\texttt{USTPU} & All Employees:  Trade, Transportation \& Utilities & $\Delta\ln$ \\
\texttt{USWTRADE} & All Employees:  Wholesale Trade & $\Delta\ln$ \\
\texttt{USGOVT} & All Employees:  Government & $\Delta\ln$ \\
\texttt{CLF16OV} & Civilian Labor Force & $\Delta\ln$ \\
\texttt{UNRATE} & Civilian Unemployment Rate & $\Delta$ \\
\texttt{UEMPMEAN} & Average Duration of Unemployment (Weeks) & $\Delta$ \\
\texttt{UEMPLT5} & Civilians Unemployed - Less Than 5 Weeks & $\Delta\ln$ \\
\texttt{UEMP5TO14} & Civilians Unemployed for 5-14 Weeks & $\Delta\ln$ \\
\texttt{UEMP15OV} & Civilians Unemployed - 15 Weeks \& Over & $\Delta\ln$ \\
\texttt{UEMP15T26} & Civilians Unemployed for 15-26 Weeks & $\Delta\ln$ \\
\texttt{UEMP27OV} & Civilians Unemployed for 27 Weeks and Over & $\Delta\ln$ \\
\texttt{CLAIMSx} & Initial Claims & $\Delta\ln$ \\
\texttt{AWHMAN} & Avg Weekly Hours :  Manufacturing & Lv \\
\texttt{AWOTMAN} & Avg Weekly Overtime Hours :  Manufacturing & $\Delta$ \\
\texttt{CES0600000007} & Avg Weekly Hours :  Goods-Producing & Lv \\
\texttt{CES0600000008} & Avg Hourly Earnings :  Goods-Producing & $\Delta^2\ln$ \\
\texttt{CES1021000001} & All Employees:  Mining and Logging:  Mining & $\Delta\ln$ \\
\texttt{CES2000000008} & Avg Hourly Earnings :  Construction & $\Delta^2\ln$ \\
\texttt{CES3000000008} & Avg Hourly Earnings :  Manufacturing & $\Delta^2\ln$ \\
\texttt{HWI}${}^{\dagger}$ & Help-Wanted Index for United States & $\Delta$ \\
\texttt{HWIURATIO}${}^{\dagger}$ & Ratio of Help Wanted/No.  Unemployed & $\Delta$ \\
\addlinespace[4pt]
\multicolumn{3}{l}{\textit{Consumption \& Inventories}} \\
\midrule
\texttt{CMRMTSPLx}${}^{\dagger}$ & Real Manu.  and Trade Industries Sales & $\Delta\ln$ \\
\texttt{RETAILx} & Retail and Food Services Sales & $\Delta\ln$ \\
\texttt{AMDMNOx} & New Orders for Durable Goods & $\Delta\ln$ \\
\texttt{AMDMUOx} & Unfilled Orders for Durable Goods & $\Delta\ln$ \\
\texttt{ANDENOx}${}^{\dagger}$ & New Orders for Nondefense Capital Goods & $\Delta\ln$ \\
\texttt{ACOGNO}${}^{\dagger}$ & New Orders for Consumer Goods & $\Delta\ln$ \\
\texttt{BUSINVx} & Total Business Inventories & $\Delta\ln$ \\
\texttt{ISRATIOx} & Total Business:  Inventories to Sales Ratio & $\Delta$ \\
\addlinespace[4pt]
\multicolumn{3}{l}{\textit{Housing}} \\
\midrule
\texttt{HOUST} & Housing Starts:  Total New Privately Owned & $\ln$ \\
\texttt{HOUSTNE} & Housing Starts, Northeast & $\ln$ \\
\texttt{HOUSTMW} & Housing Starts, Midwest & $\ln$ \\
\texttt{HOUSTS} & Housing Starts, South & $\ln$ \\
\texttt{HOUSTW} & Housing Starts, West & $\ln$ \\
\texttt{PERMIT} & New Private Housing Permits (SAAR) & $\ln$ \\
\texttt{PERMITNE} & New Private Housing Permits, Northeast (SAAR) & $\ln$ \\
\texttt{PERMITMW} & New Private Housing Permits, Midwest (SAAR) & $\ln$ \\
\texttt{PERMITS} & New Private Housing Permits, South (SAAR) & $\ln$ \\
\texttt{PERMITW} & New Private Housing Permits, West (SAAR) & $\ln$ \\
\addlinespace[4pt]
\multicolumn{3}{l}{\textit{Money \& Credit}} \\
\midrule
\texttt{M1SL} & M1 Money Stock & $\Delta^2\ln$ \\
\texttt{M2SL} & M2 Money Stock & $\Delta^2\ln$ \\
\texttt{M2REAL} & Real M2 Money Stock & $\Delta\ln$ \\
\texttt{MZMSL}${}^{\dagger}$ & MZM Money Stock & $\Delta^2\ln$ \\
\texttt{TOTRESNS} & Total Reserves of Depository Institutions & $\Delta^2\ln$ \\
\texttt{NONBORRES}${}^{\dagger}$ & Reserves Of Depository Institutions & $\Delta$ \\
\texttt{BUSLOANS} & Commercial and Industrial Loans & $\Delta^2\ln$ \\
\texttt{REALLN} & Real Estate Loans at All Commercial Banks & $\Delta^2\ln$ \\
\texttt{NONREVSL} & Total Nonrevolving Credit & $\Delta^2\ln$ \\
\texttt{INVEST} & Securities in Bank Credit at All Commercial Banks & $\Delta^2\ln$ \\
\texttt{CONSPI} & Nonrevolving consumer credit to Personal Income & $\Delta$ \\
\texttt{BOGMBASE} & Monetary Base & $\Delta^2\ln$ \\
\texttt{DTCTHFNM} & Total Consumer Loans and Leases Outstanding & $\Delta^2\ln$ \\
\texttt{DTCOLNVHFNM} & Consumer Motor Vehicle Loans Outstanding & $\Delta^2\ln$ \\
\addlinespace[4pt]
\multicolumn{3}{l}{\textit{Interest \& FX Rates}} \\
\midrule
\texttt{FEDFUNDS} & Effective Federal Funds Rate & $\Delta$ \\
\texttt{TB3MS} & 3-Month Treasury Bill: & $\Delta$ \\
\texttt{TB6MS} & 6-Month Treasury Bill: & $\Delta$ \\
\texttt{GS1} & 1-Year Treasury Rate & $\Delta$ \\
\texttt{GS5} & 5-Year Treasury Rate & $\Delta$ \\
\texttt{GS10} & 10-Year Treasury Rate & $\Delta$ \\
\texttt{AAA} & Moody's Seasoned Aaa Corporate Bond Yield & $\Delta$ \\
\texttt{BAA} & Moody's Seasoned Baa Corporate Bond Yield & $\Delta$ \\
\texttt{TB3SMFFM} & 3-Month Treasury C Minus FEDFUNDS & Lv \\
\texttt{TB6SMFFM} & 6-Month Treasury C Minus FEDFUNDS & Lv \\
\texttt{T1YFFM} & 1-Year Treasury C Minus FEDFUNDS & Lv \\
\texttt{T5YFFM} & 5-Year Treasury C Minus FEDFUNDS & Lv \\
\texttt{T10YFFM} & 10-Year Treasury C Minus FEDFUNDS & Lv \\
\texttt{CP3Mx}${}^{\dagger}$ & 3-Month AA Financial Commercial Paper Rate & $\Delta$ \\
\texttt{AAAFFM} & Moody's Aaa Corporate Bond Minus FEDFUNDS & Lv \\
\texttt{BAAFFM} & Moody's Baa Corporate Bond Minus FEDFUNDS & Lv \\
\texttt{COMPAPFFx}${}^{\dagger}$ & 3-Month Commercial Paper Minus FEDFUNDS & Lv \\
\texttt{EXCAUSx} & Canada / U.S. Foreign Exchange Rate & $\Delta\ln$ \\
\texttt{EXJPUSx} & Japan / U.S. Foreign Exchange Rate & $\Delta\ln$ \\
\texttt{EXSZUSx} & Switzerland / U.S. Foreign Exchange Rate & $\Delta\ln$ \\
\texttt{EXUSUKx} & U.S. / U.K. Foreign Exchange Rate & $\Delta\ln$ \\
\texttt{TWEXAFEGSMTHx}${}^{\dagger}$ & Trade Weighted U.S. Dollar Index & $\Delta\ln$ \\
\addlinespace[4pt]
\multicolumn{3}{l}{\textit{Prices}} \\
\midrule
\texttt{CPIAUCSL} & CPI : All Items & $\Delta^2\ln$ \\
\texttt{CPITRNSL} & CPI : Transportation & $\Delta^2\ln$ \\
\texttt{CPIAPPSL} & CPI : Apparel & $\Delta^2\ln$ \\
\texttt{CPIMEDSL} & CPI : Medical Care & $\Delta^2\ln$ \\
\texttt{CPIULFSL} & CPI : All Items Less Food & $\Delta^2\ln$ \\
\texttt{CUSR0000SA0L2} & CPI : All items less shelter & $\Delta^2\ln$ \\
\texttt{CUSR0000SA0L5} & CPI : All items less medical care & $\Delta^2\ln$ \\
\texttt{CUSR0000SAS} & CPI : Services & $\Delta^2\ln$ \\
\texttt{CUSR0000SAC} & CPI : Commodities & $\Delta^2\ln$ \\
\texttt{CUSR0000SAD} & CPI : Durables & $\Delta^2\ln$ \\
\texttt{PCEPI} & Personal Cons.  Expend.:  Chain Index & $\Delta^2\ln$ \\
\texttt{PPICMM} & PPI: Metals and metal products: & $\Delta^2\ln$ \\
\texttt{WPSID61} & PPI: Intermediate Materials & $\Delta^2\ln$ \\
\texttt{WPSID62} & PPI: Crude Materials & $\Delta^2\ln$ \\
\texttt{WPSFD49207} & PPI: Finished Goods & $\Delta^2\ln$ \\
\texttt{WPSFD49502} & PPI: Finished Consumer Goods & $\Delta^2\ln$ \\
\texttt{OILPRICEx} & Crude Oil, spliced WTI and Cushing & $\Delta^2\ln$ \\
\texttt{RPI} & Real Personal Income & $\Delta\ln$ \\
\addlinespace[4pt]
\multicolumn{3}{l}{\textit{Stock Market}} \\
\midrule
\texttt{S\&P 500} & S\&P's Common Stock Price Index: Composite & $\Delta\ln$ \\
\texttt{S\&P PE ratio}${}^{\dagger}$ & S\&P's Composite Common Stock: Price-Earnings Ratio & $\Delta\ln$ \\
\texttt{S\&P div yield}${}^{\dagger}$ & S\&P's Composite Common Stock: Dividend Yield & $\Delta$ \\
\texttt{VIXCLSx}${}^{\dagger}$ & VIX & Lv \\
\texttt{UMCSENTx}${}^{\dagger}$ & Consumer Sentiment Index & $\Delta$ \\
\addlinespace[4pt]
\end{longtable}
}%

    \begin{table}[!ht]
\centering
\caption{Variable subsets: codes, names, and FRED-MD categories}
\label{tab:variable_subsets}
\begin{threeparttable}
\begin{tabular}{lll}
\toprule
Code & Full Name & FRED-MD Group \\
\midrule
\multicolumn{3}{l}{\textit{Panel A: Medium variable set (18 series)}} \\
\midrule
\texttt{PAYEMS} & All Employees: Total Nonfarm & Labor Market \\
\texttt{INDPRO} & Industrial Production Index & Output \& Income \\
\texttt{CONSPI} & Consumer Installment Credits to Personal Income Ratio & Money \& Credit \\
\texttt{UNRATE} & Unemployment Rate & Labor Market \\
\texttt{RPI} & Real Personal Income & Prices \\
\texttt{DPCERA3M086SBEA} & Real Personal Consumption Expenditures & Output \& Income \\
\texttt{CMRMTSPLx} & Real Mfg.\ \& Trade Industries Sales & Consumption \& Inventories \\
\texttt{CUMFNS} & Capacity Utilization: Manufacturing & Output \& Income \\
\texttt{AWHMAN} & Avg.\ Weekly Hours: Manufacturing & Labor Market \\
\texttt{CES1021000001} & All Employees: Mining \& Logging & Labor Market \\
\texttt{REALLN} & Real Estate Loans, All Commercial Banks & Money \& Credit \\
\texttt{BAA} & Moody's Seasoned Baa Corporate Bond Yield & Interest \& FX Rates \\
\texttt{TB6SMFFM} & 6-Month T-Bill Minus Federal Funds Rate & Interest \& FX Rates \\
\texttt{T5YFFM} & 5-Year Treasury Rate Minus Federal Funds Rate & Interest \& FX Rates \\
\texttt{BAAFFM} & Moody's Baa Corporate Bond Minus Federal Funds Rate & Interest \& FX Rates \\
\texttt{EXCAUSx} & Canada / U.S.\ Foreign Exchange Rate & Interest \& FX Rates \\
\texttt{PCEPI} & PCE: Chain-Type Price Index & Prices \\
\texttt{PPICMM} & PPI: Commodities & Prices \\
\midrule
\multicolumn{3}{l}{\textit{Panel B: Additional series in Large variable set (+12 series)}} \\
\midrule
\texttt{CLF16OV} & Civilian Labor Force & Labor Market \\
\texttt{BUSINVx} & Real Manufacturing \& Trade Inventories & Consumption \& Inventories \\
\texttt{ISRATIOx} & Real Inventories-to-Sales Ratio & Consumption \& Inventories \\
\texttt{NONREVSL} & Total Nonrevolving Credit & Money \& Credit \\
\texttt{INVEST} & Securities in Bank Credit, All Commercial Banks & Money \& Credit \\
\texttt{DTCOLNVHFNM} & Total Consumer Loans \& Leases Outstanding & Money \& Credit \\
\texttt{AMBSL} & St.\ Louis Adjusted Monetary Base & Money \& Credit \\
\texttt{FEDFUNDS} & Effective Federal Funds Rate & Interest \& FX Rates \\
\texttt{TB6MS} & 6-Month Treasury Bill Rate & Interest \& FX Rates \\
\texttt{GS1} & 1-Year Treasury Constant Maturity Rate & Interest \& FX Rates \\
\texttt{TWEXMMTH} & Trade Weighted USD: Major Currencies & Interest \& FX Rates \\
\texttt{OILPRICEx} & Crude Oil Prices: West Texas Intermediate & Prices \\
\bottomrule
\end{tabular}
\begin{tablenotes}
\footnotesize
\item \textit{Sources:} Variable codes and descriptions follow \citet{McCracken:Ng:2016} and the FRED-MD data appendix (Federal Reserve Bank of St.~Louis, \url{https://research.stlouisfed.org/econ/mccracken/fred-databases/}). All series are transformed as specified in the FRED-MD appendix. The Large set is the union of Panel~A and Panel~B.
\end{tablenotes}
\end{threeparttable}
\end{table}

\end{appendices}
\end{document}